\documentclass[11pt,a4paper]{article}
\pdfoutput=1
\usepackage{jstyle}
\usepackage{amsmath, amsfonts, amssymb}
\usepackage{graphicx, hyperref, color, verbatim}
\usepackage[dvipsnames]{xcolor}
\usepackage{float}
\usepackage{caption}
\usepackage{subcaption}
\usepackage[bb=stix]{mathalpha}
\usepackage{tikz}
\usetikzlibrary{shapes}
\usetikzlibrary{shapes.misc}

\tikzset{
  vtx/.style={
    circle,
    draw=blue,
    fill=blue,
    inner sep=1pt
  },
  wcirc/.style={
    circle,
    draw=white,
    fill=white,
    inner sep=2pt
  },
  bcirc/.style={
    circle,
    draw=black,
    fill=black,
    inner sep=1pt
  },
  dcirc/.style={
    circle,
    draw=blue,
    fill=blue,
    inner sep=1pt
  },
  rcirc/.style={
    circle,
    draw=red,
    fill=red,
    inner sep=1pt
  },
  phi/.style={
    thick
  },
  sigma/.style={
    thick,
    dashed
  },
  vl1/.style={
    thick,
    blue
  },
  vl2/.style={
    thick,
    dashed,
    blue
  },
  valign/.style={
    baseline={([yshift=-.55ex]current bounding box.center)}
  }
}

\usepackage[LGR,T1]{fontenc}
\DeclareSymbolFont{ttgreek}{LGR}{cmtt}{m}{n}
\DeclareMathSymbol{\ttalpha}{\mathord}{ttgreek}{`a}

\makeatletter
\DeclareFontFamily{OMX}{MnSymbolE}{}
\DeclareSymbolFont{MnLargeSymbols}{OMX}{MnSymbolE}{m}{n}
\SetSymbolFont{MnLargeSymbols}{bold}{OMX}{MnSymbolE}{b}{n}
\DeclareFontShape{OMX}{MnSymbolE}{m}{n}{
    <-6>  MnSymbolE5
   <6-7>  MnSymbolE6
   <7-8>  MnSymbolE7
   <8-9>  MnSymbolE8
   <9-10> MnSymbolE9
  <10-12> MnSymbolE10
  <12->   MnSymbolE12
}{}
\DeclareFontShape{OMX}{MnSymbolE}{b}{n}{
    <-6>  MnSymbolE-Bold5
   <6-7>  MnSymbolE-Bold6
   <7-8>  MnSymbolE-Bold7
   <8-9>  MnSymbolE-Bold8
   <9-10> MnSymbolE-Bold9
  <10-12> MnSymbolE-Bold10
  <12->   MnSymbolE-Bold12
}{}

\let\llangle\@undefined
\let\rrangle\@undefined
\DeclareMathDelimiter{\llangle}{\mathopen}%
                     {MnLargeSymbols}{'164}{MnLargeSymbols}{'164}
\DeclareMathDelimiter{\rrangle}{\mathclose}%
                     {MnLargeSymbols}{'171}{MnLargeSymbols}{'171}
\makeatother

\def\I{{\mathcal{I}}}

\title{Defect Correlators from Supergravity: Finding the Missing Contact Terms}

\author[]{Qi Chen,}
\author[]{Xinan Zhou}

\affiliation[]{Kavli Institute for Theoretical Sciences, University of Chinese Academy of Sciences, Beijing 100190, China}

\abstract{We resolve a puzzle in the computation of holographic correlators using supergravity when defects are present. Previous direct supergravity calculations of defect two-point functions were found to differ from bootstrap results by a set of contact contributions. We show that the discrepancy arises from two effects: nonlinear Kaluza-Klein field redefinitions in the bulk, which induce additional quadratic couplings when applied consistently to localized sources, and off-shell terms in mixed bulk--defect vertices, which can collapse internal defect propagators. We demonstrate how these two mechanisms resolve the puzzle in two concrete examples in 4d $\mathcal{N}=4$ SYM, where we consider two-point correlators in the presence of a $\frac{1}{2}$-BPS Wilson line and on $\mathbb{RP}^4$.}

\begin{document}
\maketitle
\tableofcontents

\newpage

\section{Introduction}

Holographic correlators are basic observables in the AdS/CFT correspondence, and considerable attention in the early days of the correspondence was devoted to their computation \cite{Maldacena:1997re,Gubser:1998bc,Witten:1998qj,Lee:1998bxa,Freedman:1998tz}. The traditional approach is diagrammatic. One starts from the AdS supergravity action, or equivalently from the equations of motion, and expands around the background solution. Holographic correlators are then obtained as sums of Witten diagrams constructed from the resulting interaction vertices. Some of the first nontrivial holographic correlators were computed precisely in this way \cite{DHoker:1999pj,Arutyunov:2000py}. A new bootstrap approach was later initiated in \cite{Rastelli:2016nze,Rastelli:2017udc}, in which holographic correlators are determined by imposing symmetry and consistency conditions rather than by evaluating individual Witten diagrams. This method is substantially more efficient. Besides reproducing previously known results, the bootstrap method made it possible to obtain general holographic correlators of $\frac{1}{2}$-BPS operators with arbitrary conformal dimensions in closed form (see \cite{Bissi:2022mrs,Zhou:2026xtp} for reviews).

More recently, the bootstrap approach has been extended to CFTs with defects, and tree-level holographic correlators of local operators inserted away from the defect have been obtained in a variety of systems \cite{Barrat:2021yvp,Gimenez-Grau:2023fcy,Chen:2023yvw,Zhou:2024ekb,Chen:2025yxg,Chen:2026ium,Bianchi:2026ukq,Linardopoulos:2026mut,Chen:2026yzm}. This extension, however, also exposed a rather curious puzzle. The puzzle was first observed in \cite{Gimenez-Grau:2023fcy} for four-dimensional $\mathcal{N}=4$ SYM with an additional $\frac12$-BPS Wilson line inserted. The two-point functions of $\frac12$-BPS local operators in the presence of the $\frac{1}{2}$-BPS line defect were shown to be fully determined by the bootstrap constraints just as four-point functions in the defect-free case. But the resulting correlators are not reproduced by the naive supergravity diagrammatic calculation. More precisely, the bootstrap answer differs from the direct supergravity result by a combination of contact Witten diagrams. These additional contact contributions are required for the full correlator to satisfy the superconformal Ward identities, but their origin is not apparent from the standard expansion of the string and bulk supergravity actions. This indicates that the supergravity analysis, though seemingly standard and closely paralleling the defect-free computation, is incomplete.\footnote{A closely related mismatch was also found in \cite{Zhou:2024ekb} for two-point functions of $\frac12$-BPS operators in $\mathcal{N}=4$ SYM on $\mathbb{RP}^4$, which can be viewed as a conformal defect of dimension $-1$. We will return to this example later in the paper.} 

In this paper, we resolve the puzzle and show how the bootstrap results emerge directly from supergravity. The solution turns out to be surprisingly simple, and the missing contact contributions have two distinct origins. Both can be obscured by the standard simplifications used in reducing the bulk and defect couplings.

\vspace{0.5cm}
\noindent{\bf Nonlinear field redefinitions.} The first source of additional contact terms is the nonlinear field redefinitions needed to put bulk cubic vertices in canonical form. Let us consider for definiteness a ten-dimensional fluctuation $h_{MN}(x,y)$ where $x$, $y$ are AdS$_5$ and S$^5$ coordinates respectively. After the usual linear diagonalization of the KK spectrum, we can write the fluctuation schematically as 
\begin{equation}
    h_{MN}(x,y)=\sum_I \mathcal{K}_{MN}^I(y,\nabla_x)\widehat{s}_I(x)+\text{other KK towers}\;,
\end{equation}
where $I$ contains the KK level and representation labels, and $\mathcal{K}_{MN}^I(y,\nabla_x)\widehat{s}_I$ contains the S$^5$ harmonics, tensor structures and AdS$_5$ derivatives. We have also isolated the scalar sector $\widehat{s}_I(x)$ which are five dimensional fields. Note that $\widehat{s}_I$ are the KK fields obtained after diagonalizing the quadratic spectrum but before the nonlinear field redefinition. At second order the equation for $\widehat{s}_I$ reads schematically
\begin{equation}
    (\nabla^2-m_I^2)\widehat{s}_I=D_{IJK}\widehat{s}_J\widehat{s}_K+E_{IJK}\nabla_\mu \widehat{s}_J\nabla^\mu \widehat{s}_K+F_{IJK}\nabla_{(\mu}\nabla_{\nu)}\widehat{s}_J \nabla^{(\mu}\nabla^{\nu)}\widehat{s}_K\;.
\end{equation}
We can introduce a local nonlinear change of variables
\begin{equation}
    \widehat{s}_I=s_I+J_{IJK}s_Js_K+L_{IJK}\nabla_\mu s_J\nabla^\mu s_K+\mathcal{O}(s^3)\;,
\end{equation}
so that the equation for $s_I$ takes the form
\begin{equation}
     (\nabla^2-m_I^2)s_I=\widetilde{D}_{
     IJK
     }s_Js_K\;,
\end{equation}
with derivative terms in the cubic coupling removed. Suppose the defect in AdS has a linear coupling to the ten-dimensional field, the field redefinition then introduces quadratic terms
\begin{equation}
    \int_\Sigma d^{p+1}\sigma \mathcal{J}^{MN}(\sigma)h_{MN}(X(\sigma))\supset  \int_\Sigma d^{p+1}\sigma \mathcal{J}^{MN}\mathcal{K}^I_{MN}\widehat{s}_I=S_1[s]+S_2[s,s]+\ldots\;,
\end{equation}
where $S_1[s]$ is linear in $s_I$ while $S_2[s,s]$ is quadratic. The key point, and a potential source of confusion, is that although the nonlinear field redefinition $\widehat s_I\to s_I$ may appear to be merely a harmless renaming of variables when one considers only the bulk cubic vertices, it is not innocuous in the presence of the defect.

\vspace{0.5cm}
\noindent{\bf Off-shell terms can collapse propagators.} There is a separate way to get contact terms. Let us consider a bulk-defect coupling of the schematic form
\begin{equation}
\int_\Sigma \sqrt{\gamma}\,\widehat{\varphi}\,\widehat{\square}s\;,
\end{equation}
where the label $I$ is suppressed. For an on-shell vertex, we can integrate by parts and use the free equation of motion of the defect field to rewrite this as
\begin{equation}
\int_\Sigma \sqrt{\gamma}\,\widehat{m}^2\widehat{\varphi}s\;.
\end{equation}
The difference between the full and on-shell reduced vertices is therefore proportional to the equation-of-motion operator of the defect field. This distinction becomes important when $\widehat{\varphi}$ is exchanged in a defect exchange Witten diagram. Its propagator satisfies
\begin{equation}
(-\widehat{\square}+\widehat{m}^2)\widehat{G}(\widehat{x},\widehat{x}')
=\delta(\widehat{x}-\widehat{x}')\;,
\end{equation}
so the equation-of-motion term collapses the internal propagator and gives rise to a defect contact diagram.

\vspace{0.5cm}

It is clear that these two mechanisms should be distinguished. The first concerns the choice of field variables. A nonlinear change of variables used to put the bulk interactions into canonical form also modifies the localized source and induces new local vertices. The second concerns the use of on-shell identities inside exchange diagrams. Two vertices that are equivalent on-shell need not be equivalent when they are connected by an internal propagator. Both effects, however, have the same consequence and generate contact Witten diagrams which are absent in a naive computation. The general lesson for the supergravity calculation also becomes clear. We should not reduce bulk and defect couplings too aggressively to their on-shell forms, or replace AdS fields by their redefined versions without tracking the induced terms. A safe approach is to avoid unnecessary intermediate simplifications and keep the relevant contributions explicit. In the paper, we will show that such a recipe is sufficient and explain how to implement it in detail. We will explicitly demonstrate how these two mechanisms resolve the contact term puzzle in both the $\frac{1}{2}$-BPS Wilson line example and the case of $\mathcal{N}=4$ SYM on $\mathbb{RP}^4$.

The rest of the paper is organized as follows. In Section \ref{sugra:section} we collect the supergravity ingredients needed for the subsequent calculations. We focus on the nonlinear field redefinitions relevant for couplings of two primary scalar modes and on the off-shell form of the scalar-scalar-spin-two interaction. In Section \ref{wl:section} we study the $\frac{1}{2}$-BPS Wilson line case and show explicitly how the different contributions combine to give the required contact term. In Section \ref{o1:section} we revisit $\mathcal N=4$ SYM on $\mathbb{RP}^4$, where the nonlinear field-redefinition effect can be isolated especially cleanly. We conclude in Section \ref{sec:discussion} with a brief discussion of future directions. Further technical details are relegated to the appendices.

\section{Supergravity ingredients and field redefinitions}
\label{sugra:section}

In this section we review the necessary supergravity ingredients, focusing particularly on the couplings to a fundamental string and an O1-plane, which are relevant for the $\frac{1}{2}$-BPS Wilson loop and $\mathcal{N}=4$ SYM on $\mathbb{RP}^4$. We also discuss in detail the first mechanism for generating additional contact terms. The key point is that the ten-dimensional metric that couples to the localized D-brane/orientifold actions must be expressed in the KK variables used for the bulk cubic vertices. The nonlinear field redefinitions involved in this step generate additional localized couplings. In the F1 setup, we also have the second mechanism which generates extra contact terms from collapsing the defect mode propagators. This will be discussed separately in the Wilson-line section.

\subsection{Kaluza--Klein modes and the metric embedding}
\label{sugra:spectrum}

We consider type IIB supergravity on AdS$_5\times$S$^5$, with the common radius set to unity. We denote the ten-dimensional indices by $M$, $N$, and AdS and sphere indices by $\mu$, $\nu$ and $i$, $j$ respectively. In the metric and four-form sector, the equations take the form
\cite{Schwarz:1983qr,Howe:1983sra}
\begin{equation}
 R_{MN}=\frac16 F_{MPQRS}F_N{}^{PQRS}\;,\qquad
 F_{(5)}=*F_{(5)}\;,\qquad F_{(5)}=dA_{(4)}\;.
 \label{sugra:iib}
\end{equation}
The five-form has been normalized so that $\bar F_{(5)}=\operatorname{vol}_{AdS_5}+\operatorname{vol}_{S^5}$, which gives $\bar R_{\mu\nu}=-4\bar g_{\mu\nu}$ and $\bar R_{ij}=4\bar g_{ij}$.
These equations are written in Lorentzian signature with a mostly positive metric. The fluctuations around the background are defined by
\begin{equation}
 g_{MN}=\bar g_{MN}+\epsilon h_{MN}\;,\quad
 A_{(4)}=\bar A_{(4)}+\epsilon a_{(4)}\;,\qquad
 F_{(5)}=\bar F_{(5)}+\epsilon f_{(5)}\;,\quad f_{(5)}=da_{(4)}\;,
 \label{eq:perturbations}
\end{equation}
where $\epsilon$ keeps track of powers of the fluctuations. The spectrum follows from the linearized equations. The terms quadratic in $h_{MN}$ and $f_{(5)}$ in \eqref{sugra:iib} determine the cubic couplings and are composites of these same linear fluctuations. Hereafter we will suppress $\epsilon$.

Expansion in sphere harmonics, followed by diagonalization of the linearized equations, gives the KK spectrum \cite{Kim:1985ez,Arutyunov:1998hf}. 
In this paper, we will focus on four towers of fields which are relevant for the computation of the defect two-point functions. These are listed in Table~\ref{sugra:table}, where $k$ denotes the harmonic degree. The scalars $s_k$ and $t_k$ arise from the mixing of the internal metric trace with the four-form scalar, with $s_k$ dual to the $\frac{1}{2}$-BPS operator of dimension $k$. The field $\phi_k$ is an AdS scalar associated with an internal transverse-traceless tensor harmonic. The scalar towers \(s_k\) and \(t_k\) contribute to both F1 and O1 setups. The O1 calculation also requires the internal-tensor scalar \(\phi_k\), while the F1 calculation involves the AdS spin-two field \(\varphi_{\mu\nu,k}\).
\begin{table}[ht]
 \centering
 \begin{tabular}{||c|c|c|c|c||}
 \hline
 Field & $SU(4)$ representation & $m^2$ & $\Delta$ & Range\\[3pt]
 \hline\noalign{\vskip 1.5pt}
 \hline
 $s_k$ & $[0,k,0]$ & $k(k-4)$ & $k$ & $k\geq2$\\
 $t_k$ & $[0,k,0]$ & $(k+4)(k+8)$ & $k+8$ & $k\geq0$\\
 $\phi_k$ & $[2,k-2,2]$ & $k(k+4)$ & $k+4$ & $k\geq2$\\
 $\varphi_{\mu\nu,k}$ & $[0,k,0]$ & $k(k+4)$ & $k+4$ & $k\geq0$\\
 \hline
 \end{tabular}
 \caption{KK modes from the bulk supergravity perturbations.}
 \label{sugra:table}
\end{table}
The lowest harmonics require some care: $s_0,s_1$ do not correspond to physical modes, and the surviving $t_0,t_1$ modes have to be treated directly from the low-harmonic equations.

Before proceeding with the calculations, let us review the properties and conventions of the spherical harmonics used in our analysis. We denote the scalar spherical harmonics as $Y_k^I$ where $I$ is a degeneracy label. They satisfy the eigenvalue equation
\begin{equation}\label{sugra:scalarharmonic}
-\nabla_{S^5}^2Y_k^I=k(k+4)Y_k^I\;.
\end{equation}
We normalize the harmonics according to
\begin{equation}
\int_{S^5}d\mu_5\,Y_k^IY_k^J=z_k\delta^{IJ}\;,
\qquad z_k=\frac{1}{2^{k-1}(k+1)(k+2)}\;, \qquad d\mu_5=\frac{d\Omega_5}{\pi^3}\;.
\label{eq:harmonicnormalization}
\end{equation}
The overlap of three scalar harmonics $a_{123}$ is defined in the appendix. The group-theory contraction is included in $a_{123}$, and projection onto a scalar harmonic introduces the inverse norm $z_k^{-1}$. 

Let us now express the metric in terms of these modes. In what follows,  we will use hats to denote the fields obtained by linear diagonalization before performing nonlinear redefinitions. The internal trace $\pi=h_i^i$ has the harmonic components 
\begin{equation}
  \pi_k=10k\widehat s_k+10(k+4)\widehat t_k\;.
\end{equation}
At the linearized level, the mixing between the internal trace and the AdS metric is removed by introducing
\begin{equation}
  h_{\mu\nu}=h'_{\mu\nu}-\bar g_{\mu\nu}\frac{\pi}{3}\;.
\end{equation}
Expressing the metric perturbations $h_{\mu\nu}$ and $h_{ij}$ in terms of the KK modes, we obtain
\begin{equation}
\begin{aligned}
h_{\mu\nu}={}&\sum_{k,I}Y_k^I\bigg[\widehat\varphi_{\mu\nu,k}^I+\frac{4}{k+1}\left(\nabla_\mu\nabla_\nu-\frac{k(k-1)}2\bar g_{\mu\nu}\right)\widehat s_k^I\\
&\hspace{28mm}
+\frac{4}{k+3}\left(\nabla_\mu\nabla_\nu-\frac{(k+4)(k+5)}2\bar g_{\mu\nu}\right)\widehat t_k^I\bigg]\;,\\
h_{ij}={}&2\bar g_{ij}\sum_{k,I}Y_k^I\bigl[k\widehat s_k^I+(k+4)\widehat t_k^I\bigr]+\sum_{k,\mathcal I}\widehat\phi_k^{\mathcal I}Y_{ij,k}^{\mathcal I}\;.
\end{aligned}
\label{sugra:metric}
\end{equation}
 The straight F1 worldsheet couples to the AdS components. The O1-plane wraps an internal S$^2$ and couples to the sphere components which include also $\phi_k$. The mixed components have no linear coupling to either embedding and therefore is not presented here. The Hessian terms in \eqref{sugra:metric} also act on the quadratic field redefinitions introduced below and must therefore be kept.

\subsection{Scalar cubic interactions and field redefinitions}
\label{sugra:scalars}

To illustrate the first source of additional contact vertices, let us first consider the interaction of three scalar modes $s$.
The quadratic terms in \eqref{sugra:iib}, projected onto the harmonic
of $\widehat s_1$, can be organized according to the number of AdS
derivatives as
\begin{equation}
 (\Box-m_1^2)\widehat s_1=\sum_{2,3}\mathcal R_{123}\;,
\label{sugra:rawscalar}
\end{equation}
where
\begin{equation}
    \mathcal R_{123}=D_{123}\widehat{s}_2\widehat{s}_3 +E_{123}\nabla_\mu \widehat{s}_2\nabla^\mu \widehat{s}_3 +F_{123}\nabla_\mu\nabla_\nu \widehat{s}_2\nabla^\mu\nabla^\nu \widehat{s}_3\;.
\end{equation}
Here the subscripts $1,2,3$ label the individual KK modes, rather than their KK levels. We note that the explicit expression for $\mathcal{R}_{123}$ contains two- and four-derivative contributions arising from the supergravity perturbations. However, the derivative terms can be removed by a local field redefinition of the following form \cite{Arutyunov:1999en}
\begin{equation}
 \widehat s_1=s_1+\sum_{2,3}\mathcal A^s_{123}\;,\qquad \mathcal A^s_{123}=J_{123}s_2s_3+L_{123}\nabla s_2\cdot\nabla s_3\;,
 \label{sugra:scalarmap}
\end{equation}
where the sums run over the KK levels and the spherical-harmonic degeneracy labels of $s_2$ and $s_3$. The coefficients $J_{123}$ and $L_{123}$ are fixed by canceling the derivative terms in the equation of motion. To see this explicitly, let us define $q_{123}=m_2^2+m_3^2-m_1^2$. Using the free equations for $s_2$ and $s_3$, we obtain
\begin{equation}
\begin{aligned}
 (\Box-m_1^2)(s_2s_3) &=q_{123}s_2s_3+2\nabla s_2\cdot\nabla s_3\;,\\
 (\Box-m_1^2)(\nabla s_2\cdot\nabla s_3)&=2\nabla_\mu\nabla_\nu s_2\nabla^\mu\nabla^\nu s_3+(q_{123}-8)\nabla s_2\cdot\nabla s_3\;.
\end{aligned}
\label{sugra:products}
\end{equation}
where we have used $\bar R_{\mu\nu}=-4\bar g_{\mu\nu}$. The four- and two-derivative terms are then cancelled by choosing
\begin{equation}
 L_{123}=\frac12F_{123}\;,\qquad
 J_{123}=\frac12E_{123}+\frac14(8-q_{123})F_{123}\;.
 \label{sugra:scalarcoefficients}
\end{equation}
The resulting equation is
\begin{equation}
  (\Box-m_1^2)s_1=\sum_{2,3}\lambda_{123}s_2s_3\;,
\end{equation}
where 
\begin{equation}
    \lambda_{123}=D_{123}-q_{123}J_{123}\;.
\end{equation}
Equivalently, the original quadratic term can be written as
\begin{equation}
 \mathcal{R}_{123}=\lambda_{123}s_2s_3
 +(\Box-m_1^2)\mathcal{A}^s_{123}\;.
 \label{sugra:sourceidentity}
\end{equation}
Note that even when $\lambda_{123}=0$, the field redefinition $\mathcal{A}^s_{123}$ does not need to vanish and may induce a localized contact coupling.

The scalars $t$ and $\phi$ also require nonlinear field redefinitions of the form
\begin{equation}
\widehat\Psi_1=\Psi_1+\sum_{2,3}(J^\Psi_{123}s_2s_3+L^\Psi_{123}\nabla s_2\cdot\nabla s_3)\;,
\end{equation}
where $\Psi=t,\phi$. The relevant coefficients are given in Appendix \ref{app:scalarred}.

\subsection{Spin-two cubic interactions and field redefinitions}
\label{sugra:spin2}

The coupling of two scalar fields to a spin-two KK field requires additional care. As in the scalar case, nonlinear field redefinitions induce additional contact interactions in the localized action. However, a further subtlety arises from the trace and longitudinal components of the spin-two field. These vanish for an on-shell transverse-traceless state, but in general cannot be neglected in exchange diagrams.

More precisely, to remove the higher-derivative terms in the source, we need to perform the following field redefinition
\begin{equation}
\widehat\varphi_{\mu\nu,1}=\varphi_{\mu\nu,1}+\mathcal A^\varphi_{\mu\nu,1}\;,
\label{sugra:spinmap1}
\end{equation}
where
\begin{equation}
\mathcal A^\varphi_{\mu\nu,1}=2\nabla_{(\mu}\xi_{\nu),1}+\bar g_{\mu\nu}\eta_1+\sum_{2,3}\left[J^\varphi_{123}\nabla_{(\mu}s_2\nabla_{\nu)}s_3+L^\varphi_{123}\nabla_\rho\nabla_{(\mu}s_2\nabla^\rho\nabla_{\nu)}s_3\right]\;,
\label{sugra:spinmap2}
\end{equation}
Here $\xi_{\mu,1}$ and $\eta_1$ are respectively vector and scalar expressions quadratic in $s_2,s_3$, whose coefficients are fixed by the equations of motion of the spin-two field. In particular, their explicit forms for the $k_2=k_3=2$ case are given in Appendix \ref{app:spin2reduction}.

After this redefinition, the cubic interaction takes the form \cite{Arutyunov:1999fb}
\begin{equation}
\begin{aligned}
 \mathcal L_3(\varphi ss)=G_{123}\bigg[
 &\nabla^\mu s_2\nabla^\nu s_3\,\varphi_{\mu\nu,1}\\
 &-\frac12\left(\nabla s_2\cdot\nabla s_3+\frac{m_2^2+m_3^2-f_1}{2}s_2s_3\right)\varphi^\lambda_{\lambda,1}\bigg]\;,
\end{aligned}
\label{sugra:spinvertex}
\end{equation}
where $f_1=k_1(k_1+4)$. With the spherical harmonics normalized as specified above, the coupling coefficient $G_{123}$ is determined to be
\begin{equation}
G_{123}=\frac{4(\Sigma+2)(\Sigma+4)\alpha_1(\alpha_1-1)}{(k_2+1)(k_3+1)}a_{123}\;,~~~\Sigma=k_1+k_2+k_3\;,~~~\alpha_1=\frac{k_2+k_3-k_1}{2}\;,
\label{sugra:spincoupling}
\end{equation}
The trace term in \eqref{sugra:spinvertex} vanishes for an external transverse-traceless spin-two state, but must be retained in exchange diagrams. The longitudinal components of the exchanged field are likewise determined by the sourced spin-two equations of motion, as detailed in Appendix \ref{app:spin2reduction}.

Having discussed the nonlinear field redefinitions for both scalar and spin-two KK fields, we now summarize their effects on the localized action. Let $S_{\Sigma,\rm lin}^\Psi$ denote the part of the localized action that is linear in the KK field $\Psi$ which can be $s$, $t$, $\phi$, $\varphi$. The field redefinitions induce additional quadratic couplings obtained by replacing $\Psi$ in these linear terms with $\mathcal A^\Psi[s,s]$. Together with the direct quadratic couplings, we have
\begin{equation}
S_\Sigma^{\rm quad}
=S_{\Sigma,\rm direct}^{\rm quad}
+\sum_{\Psi=s,t,\phi,\varphi}
S_{\Sigma,\rm lin}^\Psi\big|_{\Psi\to\mathcal A^\Psi[s,s]}\;.
\label{sugra:localizedmap}
\end{equation}

The origin of these contact terms can also be understood from \eqref{sugra:sourceidentity}. In a bulk channel exchange Witten diagram, the term proportional to the equation-of-motion operator cancels the internal propagator, producing the same contact interaction as the field redefinition of the localized action. In the following calculations, we use the canonical bulk vertices and the field-redefined localized action. Additional contact terms arise from worldsheet exchanges in the F1 case, as will be discussed in the next section.

\section{$\frac{1}{2}$-BPS Wilson line}
\label{wl:section}

\subsection{Setup and vertices}\label{subsec:setupvertices}
We now apply the discussion of the supergravity couplings of Section~\ref{sugra:section} to two-point functions in 4d $\mathcal{N}=4$ SYM in the presence of a straight $\frac{1}{2}$-BPS Wilson line. Let us first briefly review the setup. For the purpose of illustration, we will focus on the two-point function of the stress tensor multiplet with the lowest weight $k=2$. However, the generalization to higher KK modes is straightforward. 

On the field theory side, we have a $\frac{1}{2}$-BPS Wilson line in the fundamental representation
\begin{equation}
W(\theta)=\frac{1}{N}{\rm tr}\,\mathcal P\exp\int_{-\infty}^{\infty}d\tau\,\bigl(iA(\tau)+\theta\cdot\Phi(\tau)\bigr)\;,\qquad \theta\cdot\theta=1\;,
\label{wl:operator}
\end{equation}
where the unit vector $\theta$ breaks the $SO(6)$ R-symmetry to $SO(5)$. We place the straight Wilson line along the $x^1$ direction and write $x=(x^1,\mathbf{x}_{\perp})$, where $\mathbf{x}_{\perp}=(x^2,x^3,x^4)$.  We consider $\frac{1}{2}$-BPS operators which are normalized to have unit defect-free two-point functions
\begin{equation}
\langle\mathcal O_k(x_1,u_1)\mathcal O_k(x_2,u_2)\rangle=\frac{(u_1\cdot u_2)^k}{|x_{12}|^{2k}}\;,\qquad u_1^2=u_2^2=0\;.
\end{equation}
Here the R-symmetry null vectors $u_i$ ensure that these operators are in the symmetric traceless representations of $SO(6)$, and the stress tensor case corresponds to $k=2$. We consider the leading connected contribution at large $N$ and large 't Hooft coupling. For the $22$ correlator, we can write it as
\begin{equation}
\frac{\langle W(\theta)\mathcal O_2(x_1,u_1)\mathcal O_2(x_2,u_2)\rangle_{\mathrm c}}{\langle W(\theta)\rangle}=\frac{\sqrt{\lambda}}{8N^2}\frac{(u_1\cdot\theta)^2(u_2\cdot\theta)^2}{|\mathbf{x}_{1\perp}|^2|\mathbf{x}_{2\perp}|^2}\mathcal F_{22}(\xi,\chi,\sigma)\;,
\label{wl:normalization}
\end{equation}
where the cross ratios are 
\begin{equation}
\xi=\frac{|x_{12}|^2}{|\mathbf{x}_{1\perp}||\mathbf{x}_{2\perp}|}\;,\qquad \chi=\frac{2\mathbf{x}_{1\perp}\cdot \mathbf{x}_{2\perp}}{|\mathbf{x}_{1\perp}||\mathbf{x}_{2\perp}|}\;,\qquad \sigma=\frac{u_1\cdot u_2}{(u_1\cdot\theta)(u_2\cdot\theta)}\;.
\label{wl:crossratios}
\end{equation}

On the dual supergravity side, the Wilson line is dual to a fundamental string which spans an AdS$_2$ surface in AdS$_5$ and sits at a point on S$^5$ \cite{Maldacena:1998im,Rey:1998ik}. In particular, the unit vector $\theta$ parameterizes the location of the F1 string on S$^5$. In Poincaré coordinates, its classical embedding is given by $z^r=0$ for $r=2,3,4$, with worldsheet coordinates $\hat z^\alpha=(\hat z_0,\hat z_1)$. We have the standard Nambu-Goto action
\begin{equation}
S_{\mathrm{F1}}=T_{\mathrm{F1}}\int d^2\hat z\,\sqrt{\det\gamma_{\alpha\beta}}\;,\qquad \gamma_{\alpha\beta}=g_{MN}(X)\partial_\alpha X^M\partial_\beta X^N\;,
\label{wl:stringaction}
\end{equation}
with $T_{\mathrm{F1}}=\frac{\sqrt{\lambda}}{2\pi}$.
For the classical embedding, $\bar\gamma_{\alpha\beta}d\hat z^\alpha d\hat z^\beta=(d\hat z_0^2+d\hat z_1^2)/\hat z_0^2$. We describe its transverse fluctuations by $X^r=\epsilon\hat z_0\zeta^r(\hat z)$ and sphere coordinates $X^i=\epsilon y^i(\hat z)$, with $i=1,\ldots,5$. The additional factor $\hat z_0$ converts the coordinate displacement into the field with a standard scalar kinetic term on AdS$_2$. Expanding \eqref{wl:stringaction} and integrating by parts gives
\begin{equation}
\left.S_{\mathrm{F1}}\right|_{\epsilon^2}=\frac{T_{\mathrm{F1}}}{2}\int_{AdS_2}d^2\hat z\,\sqrt{\bar\gamma}\,\bigl(\hat\nabla_\alpha \zeta^r\hat\nabla^\alpha \zeta^r+2\zeta^r\zeta^r+\hat\nabla_\alpha y^i\hat\nabla^\alpha y^i\bigr)\;.
\label{wl:freeaction}
\end{equation}
The five massless scalars $y^i$ and three scalars $\zeta^r$ of mass squared $m^2=2$ describe transverse fluctuations along S$^5$ and AdS$_5$, respectively. They are dual to the tilt and displacement operators on the Wilson line. We list them in Table~\ref{F1:table}.
\begin{table}[ht]
\centering
\begin{tabular}{||c|c|c|c|c||}
\hline
Fields &  $m^2$ & $\Delta$ & $SO(3)$ & $SO(5)_R $\\
\hline
\hline
$\zeta^r$ & 2 & 2 & vector & singlet\\
$y^i$ & 0 & 1 & singlet & vector\\
\hline
\end{tabular}
\caption{KK modes from the F1 supergravity perturbations.}
\label{F1:table}
\end{table}

We now consider the coupling of the string worldsheet to bulk fields. At linear order in the metric perturbation, the Nambu--Goto action gives
\begin{equation}
\left.S_{\mathrm{F1}}\right|_{\text{linear bulk field}}
=\frac{T_{\mathrm{F1}}}{2}
\int_{\mathrm{AdS}_2}d^2\hat z\,\sqrt{\bar\gamma}\,
P_\parallel^{\mu\nu}h_{\mu\nu}\;,
\label{wl}
\end{equation}
where $P_\parallel^{\mu\nu}=\bar\gamma^{\alpha\beta}\partial_\alpha\bar X^\mu\partial_\beta\bar X^\nu$ is the projector tangent to the classical worldsheet. Substituting the KK decomposition \eqref{sugra:metric}, we obtain the linear coupling density
\begin{equation}
\begin{aligned}
\mathcal V_{\mathrm{F1}}^{(0,1)}=\sum_{k,I}Y_k^I(\theta)\bigg[&\frac12P_\parallel^{\mu\nu}\widehat\varphi_{\mu\nu,k}^I+\frac{2}{k+1}\bigl(\hat\Box\widehat s_k^I-k(k-1)\widehat s_k^I\bigr)\\
&+\frac{2}{k+3}\bigl(\hat\Box\widehat t_k^I-(k+4)(k+5)\widehat t_k^I\bigr)\bigg]\;.
\end{aligned}
\label{wl:linearKK}
\end{equation}
Here the common prefactor $T_{\mathrm{F1}}\sqrt{\bar\gamma}$ is suppressed, and we denote by $V_{\mathrm{F1}}^{(m,n)}$ a coupling
density of order $m$ in worldsheet fluctuations and order $n$ in bulk fields. The worldsheet Laplacian $\hat\Box$ arises from the pullback of the AdS covariant Hessians. For separated external insertions, these terms integrate to vanishing boundary contributions.

The nonlinear KK field redefinitions discussed in Section \ref{sugra:section} induce additional quadratic couplings when substituted into the linear vertex \eqref{wl:linearKK}. For example, the scalar
redefinition \eqref{sugra:scalarmap} gives
\begin{equation}
\mathcal V_{\mathrm{F1,\,f.r.}}^{(0,2)}\big|_s=\sum_{1,2,3}\frac{2Y_{k_1}^{I_1}(\theta)}{k_1+1}\bigl[\hat\Box-k_1(k_1-1)\bigr]\bigl(J_{123}s_2s_3+L_{123}\nabla_\mu s_2\nabla^\mu s_3\bigr)\;.
\label{wl:inducedscalar}
\end{equation}
The $t$ and spin-two towers contribute analogously. Their explicit contributions are evaluated in Appendix \ref{appWL:details}. The direct quadratic coupling follows from the
expansion of the Nambu--Goto action at
fixed embedding
\begin{equation}
\mathcal V_{\mathrm{F1,\, dir}}^{(0,2)}=\frac18\bigl(P_\parallel^{\mu\nu}h_{\mu\nu}\bigr)^2-\frac14P_\parallel^{\mu\rho}P_\parallel^{\nu\sigma}h_{\mu\nu}h_{\rho\sigma}\;.
\label{wl:determinant}
\end{equation}

A second source of contact terms arises from the off-shell mixed bulk-worldsheet vertices. Varying the string embedding in the linear metric coupling and integrating by parts, we get for $k=2$
\begin{equation}
\mathcal{V}_{\mathrm{F1}}^{(1,1)}=\frac23\sum_I\left[y^i\partial_iY_2^I(\theta)(\hat{\Box}-2)s_2^I-\zeta^rY_2^I(\theta)\hat{\Box}\bigl(\hat z_0\partial_rs_2^I\bigr)\right].
\label{wl:mixed}
\end{equation}
The normal derivative is taken before restricting the bulk field to the string worldsheet. To make the off-shell contributions explicit, let us denote $f_r^I=\hat z_0\partial_r s_2^I$. Integrating by parts on the worldsheet, we obtain
\begin{equation}
\begin{aligned}
y^i(\hat\Box-2)s_2^I &\simeq -s_2^I(-\hat\Box)y^i-2s_2^Iy^i\;,\\
-\zeta^r\hat\Box f_r^I &\simeq f_r^I(-\hat\Box+2)\zeta^r-2f_r^I\zeta^r\;,\label{worldsheetibp}
\end{aligned}
\end{equation}
where $\simeq$ denotes equality up to worldsheet total derivatives. This coupling contains the equation-of-motion operators of the worldsheet fields, whose Green functions obey
\begin{equation}
-\hat\Box_{\hat z}\widehat G_1(\hat z,\hat w)=\delta_{AdS_2}(\hat z,\hat w)\;,\qquad (-\hat\Box_{\hat z}+2)\widehat G_2(\hat z,\hat w)=\delta_{AdS_2}(\hat z,\hat w)\;.
\label{wl:Green}
\end{equation}
The equation-of-motion operators in these vertices cancel internal worldsheet propagators through (\ref{wl:Green}), thereby generating local contact terms. These contributions would be missed if the mixed vertices were reduced using the free worldsheet equations of motion before evaluating the exchange diagrams.

The total contact contribution therefore comes from three sources. In addition to the direct quadratic vertex, we also have the terms induced by nonlinear KK field redefinitions, and the local contributions from collapsed worldsheet propagators. We now evaluate their sum for the $22$ correlator.

\subsection{Contact contribution and comparison with bootstrap}
\begin{figure}[tb]
\centering
\begin{tikzpicture}[
  x=1cm,y=1cm,
  line cap=round,line join=round,
  boundary/.style={black,line width=.9pt},
  external/.style={
    gray!75!black,dashed,line width=.9pt,
    dash pattern=on 3.5pt off 2.6pt
  },
  worldsheet/.style={cyan!23,line width=4.5pt},
  worldline/.style={cyan!55!blue,line width=.9pt},
  bulkvertex/.style={
    circle,fill=cyan!60!blue,inner sep=2.0pt
  },
  defectprop/.style={
    gray!85!black,dashed,line width=1.05pt,
    dash pattern=on 3pt off 2.2pt
  },
  every node/.style={font=\small}
]
\foreach \dx in {0,4.70,9.40}{
  \begin{scope}[shift={(\dx,0)}]
    \begin{scope}
      \clip (0,0) circle[radius=1.55];
      \draw[worldsheet]
        (-1.15,-1.04) ..
        controls (-.62,-.53) and (.62,-.53)
        .. (1.15,-1.04);
      \draw[worldline]
        (-1.15,-1.04) ..
        controls (-.62,-.53) and (.62,-.53)
        .. (1.15,-1.04);
    \end{scope}
    \draw[boundary] (0,0) circle[radius=1.55];
    \node at (-1.17,1.43) {$x_1$};
    \node at ( 1.17,1.43) {$x_2$};
  \end{scope}
}
\begin{scope}
  \draw[external] (-.80,1.327) -- (0,-.6575);
  \draw[external] ( .80,1.327) -- (0,-.6575);
  \node[bulkvertex] at (0,-.6575) {};
  \node at (0,-1.92) {(a)};
\end{scope}
\begin{scope}[shift={(4.70,0)}]
  \draw[external] (-.80,1.327) -- (0,.22);
  \draw[external] ( .80,1.327) -- (0,.22);
  \draw[external] (0,.22) -- (0,-.6575);
  \node[bulkvertex] at (0,.22) {};
  \node[bulkvertex] at (0,-.6575) {};
  \node at (0,-1.92) {(b)};
\end{scope}
\begin{scope}[shift={(9.40,0)}]
  \draw[external] (-.80,1.327) -- (-.607,-.740);
  \draw[external] ( .80,1.327) -- ( .607,-.740);
  \draw[defectprop]
    (-.607,-.740) ..
    controls (-.36,-.30) and (.36,-.30)
    .. (.607,-.740);
  \node[bulkvertex] at (-.607,-.740) {};
  \node[bulkvertex] at ( .607,-.740) {};
  \node at (0,-1.92) {(c)};
\end{scope}
\end{tikzpicture}
\caption{Three types of defect Witten diagrams for Wilson loop: (a) contact, (b) bulk exchange and (c) defect exchange. The blue curve represents the fundamental string worldsheet.}
\label{wl:diagrams}
\end{figure}
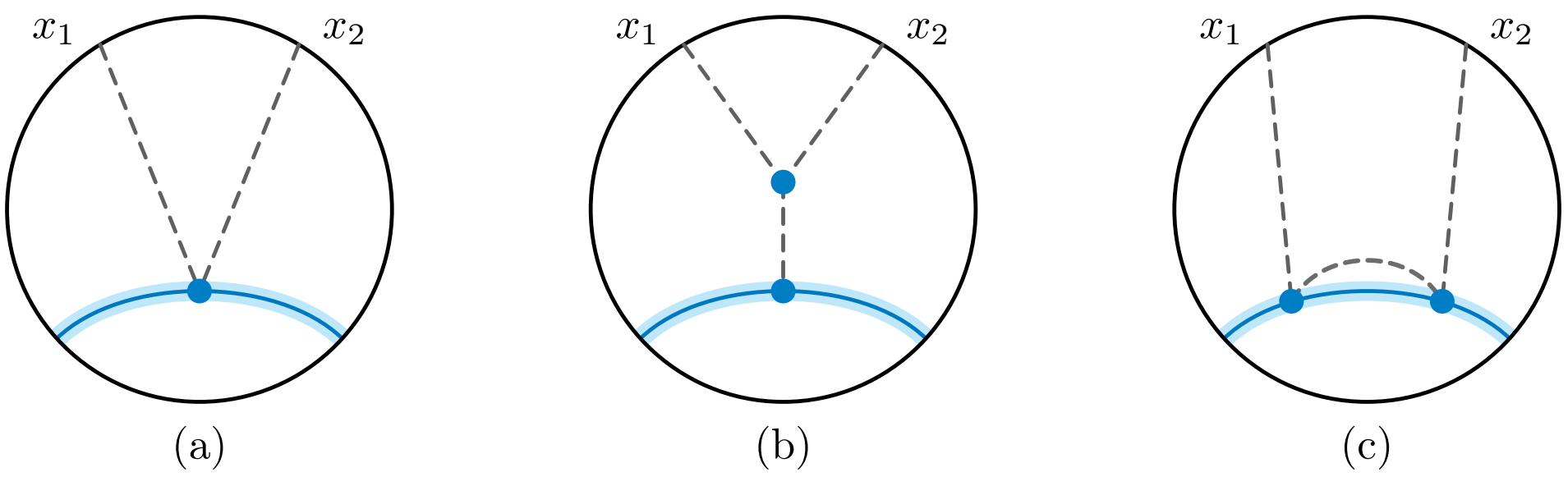

 In the supergravity approximation, the leading connected defect two-point function is given as a sum of three types of tree-level Witten diagrams. These are the exchange Witten diagrams in the bulk and defect channels, and the contact Witten diagrams, see Figure \ref{wl:diagrams}. Our focus in this subsection is the contact contribution since that was the source of disagreement. It is useful to first recall some basics. We use the bulk-to-boundary propagator 
\begin{equation}
  K_\Delta(z,x)=\left(\frac{z_0}{(z_0^2+|\vec z-x|^2)}\right)^\Delta\;,
\end{equation}
and define the basic contact Witten diagram by
\begin{equation}
\frac{C_{\Delta_1\Delta_2}(\xi,\chi)}{|\mathbf{x}_{1\perp}|^{\Delta_1}|\mathbf{x}_{2\perp}|^{\Delta_2}}=\int_{AdS_2}\frac{d^2\hat z}{\hat z_0^2}\,K_{\Delta_1}(\hat z,x_1)K_{\Delta_2}(\hat z,x_2)\;.
\label{wl:contact}
\end{equation}
Contact Witten diagrams with derivatives also reduce to $C_{\Delta_1\Delta_2}$, but with shifted dimensions. For example, we have
\begin{equation}
|\mathbf{x}_{1\perp}|^2|\mathbf{x}_{2\perp}|^2\int_{AdS_2}\sqrt{\bar\gamma}\,\nabla_\mu K_2(z,x_1)\nabla^\mu K_2(z,x_2)=4C_{22}-8\xi C_{33}\;,
\end{equation}
and a contraction tangent to the string instead produces $4C_{22}-8(\xi+\chi)C_{33}$. Further derivative contractions are collected in Appendix \ref{appWL:details}. The contact Witten diagrams can be evaluated in a closed form as ${}_2F_1$ functions \cite{Rastelli:2017ecj,Gimenez-Grau:2023fcy}
\begin{equation}
C_{\Delta_1\Delta_2}(\xi,\chi)=\frac{\pi}{2^{\Delta_1+\Delta_2}}
\frac{\Gamma\left(\frac{\Delta_1+\Delta_2-1}{2}\right)}{\Gamma\left(\frac{\Delta_1+\Delta_2+1}{2}\right)}{}_2F_1\left(\Delta_1,\Delta_2;
\frac{\Delta_1+\Delta_2+1}{2};\frac{2-\xi-\chi}{4}\right)\;.
\label{wl:contact-hypergeom}
\end{equation}

We now evaluate the contact contributions from the vertices obtained in Section \ref{subsec:setupvertices}. The direct quadratic term follows by substituting the linearized metric (\ref{sugra:metric}) into (\ref{wl:determinant}).
The remaining contributions come from the nonlinear KK field redefinitions and the local parts of the worldsheet exchanges, obtained using (\ref{worldsheetibp}) and (\ref{wl:Green}) respectively. The harmonic contractions and derivative integrals are given in Appendix \ref{appWL:details}. Using (\ref{wl:contact-hypergeom}) to simplify their sum, we find
\begin{equation}
\begin{aligned}
F_{22,\mathrm{direct}}={}&\frac{1}{\pi}\big[-24C_{22}+32(\xi+\chi)C_{33}+1152C_{44}\big]\;,\\
F_{22,\mathrm{extra}}={}&\frac{1}{\pi}\big[(32-16\sigma)C_{22}-32(\xi+\chi)C_{33}-1152C_{44}\big]\;.
\end{aligned}
\label{wl:contact-split}
\end{equation}
Here $F_{22,\mathrm{extra}}$ includes the contributions from the nonlinear KK field redefinitions and the collapsed worldsheet propagators. The derivative contact terms cancel between the two contributions, leaving
\begin{equation}
F_{22,\mathrm{contact}}=\frac{8}{\pi}(1-2\sigma)C_{22}\;.
\label{wl:contact-total}
\end{equation}

The bulk and defect exchange diagrams have already been evaluated in \cite{Gimenez-Grau:2023fcy}. Adding them to the contact contribution gives the complete correlator
\begin{equation}
\begin{aligned}
F_{22}(\xi,\chi,\sigma)
=\frac8\pi\bigg[
&8\sigma\left(1-\frac{\sigma}{6}\right)
\mathcal E_{22}^{2,0}
-2\sigma^2\mathcal E_{22}^{4,2}\\
&+4(\sigma-1)\widehat{\mathcal E}_{22}^{1,0}
+\widehat{\mathcal E}_{22}^{2,1}
+(1-2\sigma)C_{22}
\bigg]\;.
\end{aligned}
\end{equation}
Here $\mathcal E_{22}^{\Delta,\ell}$ and
$\widehat{\mathcal E}_{22}^{\widehat\Delta,s}$
denote the bulk and defect channel exchange Witten diagrams respectively, with their normalizations specified in Appendix \ref{appWL:details}. For the spin-two exchange we use the contraction with the tangent projector $P_\parallel^{\mu\nu}$. Its relation to the convention of \cite{Gimenez-Grau:2023fcy} is given in Appendix \ref{appWL:details}.

The supergravity result now agrees with the bootstrap correlator of \cite{Gimenez-Grau:2023fcy} and satisfies the superconformal Ward identities. In particular, the contact contribution required by the bootstrap is now correctly reproduced by the total quadratic couplings and the local terms from worldsheet exchanges. Although here we only explicitly show the 22 correlator as an example, we have also used the same method to compute the higher weight case of 33 and found the results agree with the bootstrap one.

\section{$\mathcal{N}=4$ SYM on $\mathbb{RP}^4$}
\label{o1:section}

\subsection{Setup and vertices}
Let us now consider $\mathcal N=4$ SYM on $\mathbb{RP}^4$. We also start with reviewing the field theory setup. Here $\mathbb{RP}^4$ is constructed as $\mathbb{R}^4$ quotiented by the conformal inversion
\begin{equation}
    \mathcal{I}:\quad x^\mu\to x'^\mu=-\frac{x^\mu}{x^2}\;.
\end{equation}
We identify the fields in the Lagrangian at $x$ and $x'$. This is accompanied by conjugating the conformal version $\mathbb{Z}_2$ with the charge conjugation and another $\mathbb{Z}_2$ acting on R-symmetry. For example, for the six scalars in $\mathcal{N}=4$ SYM we have
\begin{equation}
    \Phi^a_I(x'){\rm t}_a=-  \bar{\Phi}^a_I(x){\rm t}_a^{\rm T}\;,
\end{equation}
where $t_a$ is the adjoint color matrix and 
\begin{equation}
    \bar{\Phi}=-\mathsf{R}\Phi\;,
\end{equation}
with $\mathsf{R}={\rm diag}(1,1,1,-1,-1,-1)$.
This setup preserves half of the superconformal symmetry \cite{Wang:2020jgh,Caetano:2022mus}. 
For $\frac{1}{2}$-BPS operators, we normalize them in the same way as in the previous section. On $\mathbb{RP}^4$ their two-point functions can be written as
\begin{equation}
   \llangle \mathcal{O}_{k_1} \mathcal{O}_{k_2} \rrangle =\frac{(u_1\cdot \bar{u}_1)^{\frac{k_1}{2}}(u_2\cdot \bar{u}_2)^{\frac{k_2}{2}}}{(1+x_1^2)^{k_1}(1+x_2^2)^{k_2}}\mathcal{G}_{k_1k_2}(\eta,\sigma,\bar{\sigma})\;,\label{o1:correlator}
\end{equation}
where $\bar{u}=\mathsf{R} u$ and the cross ratios are defined as 
\begin{equation}
    \eta=\frac{x_{12}^2}{(1+x_1^2)(1+x_2^2)}\;,
\end{equation}
\begin{equation}
    \sigma=\frac{u_1\cdot u_2}{(u_1\cdot \bar{u}_1)^{\frac{1}{2}}(u_2\cdot \bar{u}_2)^{\frac{1}{2}}}\;,\quad \bar{\sigma}=\frac{u_1\cdot \bar{u}_2}{(u_1\cdot \bar{u}_1)^{\frac{1}{2}}(u_2\cdot \bar{u}_2)^{\frac{1}{2}}}\;.
\end{equation}

On the dual side, this setup corresponds to taking a $\mathbb{Z}_2$ quotient of AdS$_5\times$S$^5$ which acts on the Poincar\'e coordinates of AdS$_5$ as
\begin{equation}
    (z_0,z_\mu)\to \left(\frac{z_0}{z_0^2+(z_\mu)^2},-\frac{z_\mu}{z_0^2+(z_\mu)^2}\right)\;,
\end{equation}
and on the S$^5$ embedding coordinates $n^A$, $n^An^A=1$, as 
\begin{equation}
    n \to \mathsf{R} n\;.
\end{equation}

The quotient has a fixed locus
$\Sigma_{\mathrm{O1}}=\{z_\star\}\times {\rm S}^2$, where
$z_\star=(1,\vec 0)$ and
${\rm S}^2=\{n\in {\rm S}^5:\mathsf{R}n=n\}$.
An O1-plane wraps this S$^2$ and is localized at $z_\star$ in AdS$_5$~\cite{Wang:2020jgh,Caetano:2022mus,Zhou:2024ekb,Caetano:2026jee}. Its action in the sector relevant to our calculation is
\begin{equation}
\begin{aligned}
S_{\mathrm{O1}}&=-T_{\mathrm{O1}}\int_{S^2}d^2\omega\,\sqrt{\det\gamma_{mn}}\;,\qquad\gamma_{mn}&=g_{ij}(z_\star,\bar X(\omega))\partial_m\bar X^i\partial_n\bar X^j\;.
\end{aligned}
\label{o1:action}
\end{equation}
Here $\omega^m$ are coordinates on the wrapped S$^2$, $\bar X^i(\omega)$ describes its embedding in S$^5$, and $T_{\mathrm{O1}}$ includes the constant dilaton and the normalization of the orientifold source. The metric entering (\ref{o1:action}) is the physical ten-dimensional metric.

Introducing the tangent projector
$P_{S^2}^{ij}=
\bar\gamma^{mn}\partial_m\bar X^i
\partial_n\bar X^j$, where $\bar\gamma_{mn}$ is the induced metric obtained by replacing $g_{ij}$ with
the background metric $\bar g_{ij}$, we expand (\ref{o1:action}) to linear order in the bulk fields and substitute the KK decomposition
\eqref{sugra:metric}. This gives
\begin{equation}
\begin{aligned}
\left.S_{\mathrm{O1}}\right|_{\mathrm{linear}}=-T_{\mathrm{O1}}\int_{S^2}d\Omega_2\bigg[&2\sum_{k,I}Y_k^I(n)\bigl(k\widehat s_k^I+(k+4)\widehat t_k^I\bigr)+\frac12\sum_{k,\mathcal I}P_{S^2}^{ij}Y_{ij,k}^{\mathcal I}(n)\widehat\phi_k^{\mathcal I}\bigg]_{z=z_\star}\;,
\end{aligned}
\label{o1:linear}
\end{equation}
where $d\Omega_2=\sqrt{\bar{\gamma}}d^2\omega$ is the volume element of the unit S$^2$, and the harmonics are normalized as in (\ref{eq:harmonicnormalization}).

For two external scalars $s_p$, the linear perturbation of the induced metric is proportional to $\bar\gamma_{mn}$. Since the O1 worldvolume is two-dimensional, the direct quadratic contribution from expanding the determinant vanishes. The quadratic contact coupling therefore comes entirely from the nonlinear KK field redefinitions. Substituting the redefinitions of Section \ref{sugra:section} into the linear coupling (\ref{o1:action}), we obtain
\begin{equation}
\begin{aligned}
\left.S_{\mathrm{O1}}\right|_{s_ps_p}=-T_{\mathrm{O1}}\int_{S^2}d\Omega_2\bigg[&2\sum_{k,I}Y_k^I\bigl(k\mathcal A_k^{s,I}[s_p,s_p]+(k+4)\mathcal A_k^{t,I}[s_p,s_p]\bigr)\\
&+\frac12\sum_{k,\mathcal I}P_{S^2}^{ij}Y_{ij,k}^{\mathcal I}\mathcal A_k^{\phi,\mathcal I}[s_p,s_p]\bigg]_{z=z_\star}.
\end{aligned}
\label{o1:induced}
\end{equation}
Here $\mathcal{A}^{s,I}_k$, $\mathcal{A}^{t,I}_k$, and
$\mathcal{A}^{\phi,I}_k$ are the quadratic composites defined in Section \ref{sugra:section}, with coefficients given in Appendix \ref{app:scalarred}. In particular, the exceptional $t_0$ mode and the internal tensor modes must be included even when the corresponding canonical cubic couplings vanish.

It is important to note that unlike the Wilson line case, the O1-plane is fixed by the quotient and there are no worldvolume fluctuation that can appear in the exchanges. We now evaluate the contact couplings (\ref{o1:induced}) for the $22$ and $33$ correlators.

\subsection{Contact contributions and comparison with bootstrap}
\begin{figure}[tb]
\centering
\begin{tikzpicture}[x=1cm,y=1cm,font=\small,line cap=round,line join=round]
\begin{scope}
\draw[black,line width=0.75pt] (0,0) circle (1.8);
\draw[gray,line width=0.65pt,dash pattern=on 2.5pt off 2.5pt] (140:1.8)--(0,0)--(40:1.8);
\fill[black] (140:1.8) circle (0.05);
\fill[black] (40:1.8) circle (0.05);
\node[above=4pt,align=center] at (140:2) {$x_1$};
\node[above=4pt,align=center] at (40:2) {$x_2$};
\fill[fill={rgb,255:red,230;green,159;blue,0}] (0,0) circle (0.105);
\node[below=5pt] at (0,0) {$z_\star$};
\node at (0,-2.3) {(a) Contact};
\end{scope}
\begin{scope}[xshift=5.05cm]
\draw[black,line width=0.75pt] (0,0) circle (1.8);
\draw[gray,line width=0.65pt,dash pattern=on 2.5pt off 2.5pt] (140:1.8)--(0,0.7)--(40:1.8);
\draw[gray,line width=0.65pt] (0,0.7)--(0,0);
\fill[black] (140:1.8) circle (0.05);
\fill[black] (40:1.8) circle (0.05);
\node[above=4pt,align=center] at (140:2) {$x_1$};
\node[above=4pt,align=center] at (40:2) {$x_2$};
\fill[fill={rgb,255:red,0;green,114;blue,178}] (0,0.7) circle (0.105);
\node[left=5pt] at (0,0.57) {$z$};
\node[right=5pt] at (0,0.35) {$\Psi_k$};
\fill[fill={rgb,255:red,230;green,159;blue,0}] (0,0) circle (0.105);
\node[below=5pt] at (0,0) {$z_\star$};
\node at (0,-2.3) {(b) Direct exchange};
\end{scope}
\begin{scope}[xshift=10.1cm]
\draw[black,line width=0.75pt] (0,0) circle (1.8);
\draw[gray,line width=0.65pt,dash pattern=on 2.5pt off 2.5pt] (140:1.8)--(0,0.7)--(40:1.8);
\draw[gray,line width=0.65pt] (0,0.7)--(0,0);
\fill[black] (140:1.8) circle (0.05);
\fill[black] (40:1.8) circle (0.05);
\node[above=4pt,align=center] at (140:2) {$x_1$};
\node[above=4pt,align=center] at (40:2) {$-x_2/|x_2|^2$};
\fill[fill={rgb,255:red,0;green,114;blue,178}] (0,0.7) circle (0.105);
\node[left=5pt] at (0,0.57) {$z$};
\node[right=5pt] at (0,0.35) {$\Psi_k$};
\fill[fill={rgb,255:red,230;green,159;blue,0}] (0,0) circle (0.105);
\node[below=5pt] at (0,0) {$z_\star$};
\node at (0,-2.3) {(c) Image exchange};
\end{scope}
\end{tikzpicture}
\caption{Witten diagrams for $\mathcal{N}=4$ SYM on $\mathbb{RP}^4$: (a) contact, (b) direct exchange, and (c) image exchange. The orange point represents the O1 at the fixed point $z_\star$ in AdS$_5$. Here the exchange Witten diagram is computed by the method of images. Panel (c) is the image of (b) by replacing the second boundary point $x_2$ with $-x_2/|x_2|^2$.}
\label{o1:diagrams}
\end{figure}
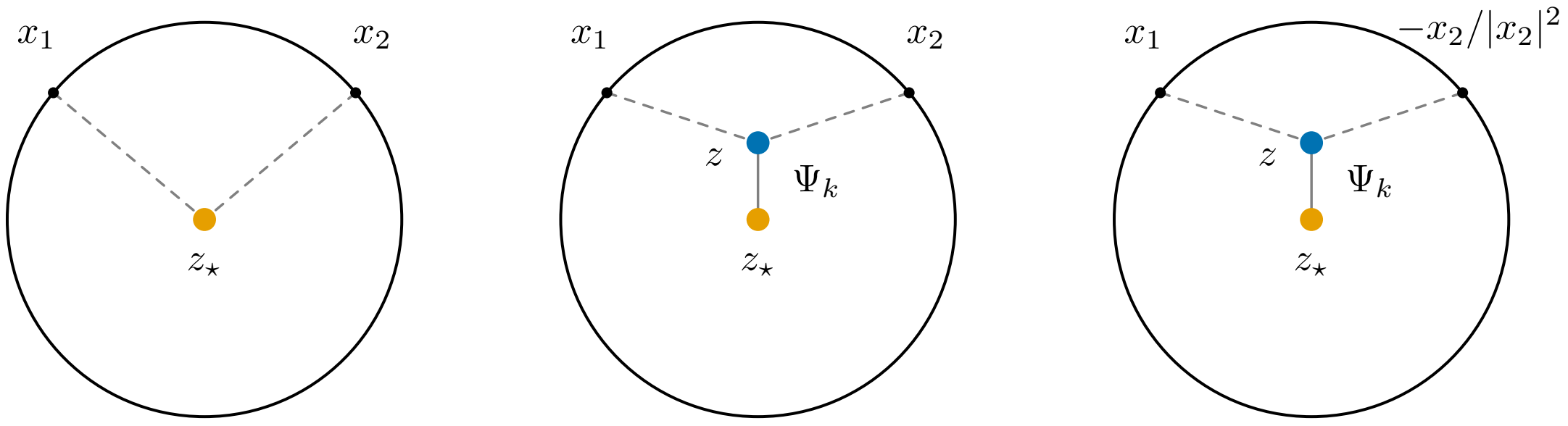

The Witten diagrams in this quotient setup can be computed via the method of images \cite{Giombi:2020xah}, and the relevant Witten diagrams are listed in Figure \ref{o1:diagrams}. In particular, the normalized bulk-to-boundary propagator is  
\begin{equation}
K_p^{\text{O1}}(z,x)=\frac{2^{p/2-2}(p+1)}{N\sqrt p}\left[K_p(z,x)(u\cdot n)^p+K_p(\iota z,x)(\bar{u}\cdot n)^p\right]\;.
\label{o1:quotientwave}
\end{equation}
Here $K_p$ denotes the bulk-to-boundary propagator defined in Section~\ref{wl:section}. The two terms are the direct contribution and its $\mathbb Z_2$ image respectively. At the O1 locus, we note $\mathcal{I} z_\star=z_\star$ and $\mathsf{R}n=n$. Therefore, the direct and image contributions coincide. Each bulk-to-boundary propagator therefore gives twice the direct contribution when evaluated at the fixed locus. Note that contact Witten diagrams are particularly simple. They are just a product of bulk-to-boundary propagators because there is no integral to perform at $z_\star$.

Let us first consider the $22$ correlator. Specializing the quadratic coupling \eqref{o1:induced} to two external $s_2$ fields, and performing the harmonic projections and integrals over the wrapped S$^2$, we obtain
\begin{equation}
\left.S_{\mathrm{O1}}\right|_{s_2s_2}=-\frac{2T_{\mathrm{O1}}}{\pi}\left[\sum_I s_2^I(z_\star)\int_{S^2}d\Omega_2\,Y_2^I(n)\right]^2.
\label{o1:quadratic}
\end{equation}
Inserting the bulk-to-boundary propagator
\eqref{o1:quotientwave}, we use
$K_2(z_\star,x)=(1+x^2)^{-2}$.
The wrapped harmonic integrals are evaluated in Appendix \ref{appO1:details}. After extracting the kinematic
prefactor as in \eqref{o1:correlator}, we obtain
\begin{equation}
\mathcal G_{22,\mathrm{contact}}
=\frac{8\pi T_{\mathrm{O1}}}{N^2}\;.
\label{o1:contact22}
\end{equation}
The bulk exchange contribution was already obtained in \cite{Zhou:2024ekb}. Including the image contribution, it reads
\begin{equation}
\mathcal G_{22,\mathrm{exchange}}
=\frac{16\pi T_{\mathrm{O1}}}{N^2}
\frac{\sigma\bar\sigma}{\eta(1-\eta)}\;.
\end{equation}
Adding the contact contribution, we obtain
\begin{equation}
\mathcal{G}_{22}(\eta,\sigma,\bar\sigma)=\frac{4\pi T_{\mathrm{O1}}}{N^2}\left[\frac{4\sigma\bar\sigma}{\eta(1-\eta)}+2\right].
\end{equation}
The contact term agrees precisely with the bootstrap result of \cite{Zhou:2024ekb}, with the additive bootstrap ambiguity fixed to zero.

We next consider the $33$ correlator. In this case,
the quadratic O1 coupling receives contributions
from scalar harmonics of degrees $0,2,4,6$ and
tensor harmonics of degrees $2,4$.
After summing these contributions, the terms
containing AdS derivatives cancel, and we obtain
\begin{equation}
\mathcal{G}_{33,\mathrm{contact}}=\frac{16\pi T_{\mathrm{O1}}}{N^2}(\sigma+\bar\sigma)\;.
\label{o1:contact33}
\end{equation}
The details are given in Appendix \ref{appO1:details}. Adding the $s_2$ and $s_4$ bulk exchanges and their $\mathbb Z_2$ images, we obtain
\begin{equation}
\begin{aligned}
\mathcal{G}_{33}=\frac{4\pi T_{\mathrm{O1}}}{N^2}
\bigg[
&4(\sigma+\bar\sigma)+6\sigma\bar\sigma
\left(
\frac{(1+\eta)\sigma}{\eta^2}
+\frac{(2-\eta)\bar\sigma}{(1-\eta)^2}
\right)\\
&+2\left(
\frac{\sigma(2-\sigma^2+4\bar\sigma^2)}{\eta}
+\frac{\bar\sigma(2-\bar\sigma^2+4\sigma^2)}
{1-\eta}
\right)
\bigg]\;.
\end{aligned}
\label{o1:full33}
\end{equation}
This agrees with the bootstrap result of \cite{Zhou:2024ekb}. Unlike the $22$ correlator, superconformal Ward identities do not allow any additive contact ambiguity for $33$. The supergravity calculation reproduces the required contact contribution without any additional terms. Moreover, we have checked the $24$ and $44$ correlators and found they are in agreement with the bootstrap results.

\section{Discussion}\label{sec:discussion}

In this paper, we resolved a mismatch between direct supergravity calculations and bootstrap results for holographic defect two-point functions. The solution turned out to be remarkably simple. We identified two sources of contact contributions that can be easily overlooked. The first arises from nonlinear Kaluza--Klein field redefinitions, which induce additional quadratic couplings in the localized action. The second arises from off-shell terms in mixed bulk-defect vertices, which can collapse internal defect propagators. We demonstrated these mechanisms in two examples of defects in 4d $\mathcal{N}=4$ SYM, namely the $\frac{1}{2}$-BPS Wilson line and the theory on $\mathbb{RP}^4$. Using the correct prescription, the results which we computed from supergravity agree with those obtained using bootstrap methods \cite{Gimenez-Grau:2023fcy,Zhou:2024ekb}. This resolves the previously observed discrepancies.

Although we have considered only two setups in detail, the same mechanisms should apply more generally. One natural extension is to consider $\frac{1}{2}$-BPS surface defects in the 6d $(2,0)$ theory, dual to an M2-brane wrapping AdS$_3$ in AdS$_7\times$S$^4$. This setup is closely related to the examples studied here, and the defect two-point function of $\frac{1}{2}$-BPS local operators have already been computed using bootstrap methods \cite{Meneghelli:2022gps,Chen:2023yvw}. Another particularly interesting setup is four-point functions in $\mathcal{N}=4$ SYM with two light $\frac{1}{2}$-BPS operators and two heavy giant gravitons. These correlators can be viewed as two-point functions in the presence of a zero-dimensional defect represented by the giant graviton pair. For maximal giant gravitons, these correlators have been computed for arbitrary light operator dimensions using bootstrap techniques \cite{Chen:2025yxg,Chen:2026ium}. Recently, the result was extended to identical heavy operators which can be either non-maximal or dual giant gravitons and can have arbitrary allowed sizes \cite{Chen:2026yzm}. A special feature of this setup is the $U(1)$ collective coordinate of the classical giant graviton solution which describes how the dual D3-brane is embedded in AdS$_5\times$S$^5$. The corresponding orbital average is needed to project onto quantum states of definite $U(1)$ charge. It would be interesting to implement this projection together with the contact interactions discussed here and reproduce these correlators directly from the D3-brane action.

Beyond reproducing the known results, it would also be interesting to compute defect correlators in new setups. A good target is the probe D3-D5 system, where the D5-brane wraps AdS$_4\times$S$^2$ and describes a co-dimension one interface \cite{Karch:2000gx,Karch:2001cw,DeWolfe:2001pq}. Of particular interest is the configuration with magnetic flux on the wrapped S$^2$ which corresponds to D3-branes ending on the D5-brane. The flux produces a tilted AdS$_4$ embedding inside AdS$_5$. Due to the simpler co-dimention one kinematics, this is also an attractive setup for studying higher-point correlators. In particular, the basic Witten diagrams for interface three-point functions have been computed in \cite{Chen:2023oax}, providing useful ingredients for a direct supergravity calculation.

\acknowledgments

This work is supported by the NSFC Grant No. 12275273 and funds from the Chinese Academy of Sciences. Generative AI was used to assist in checking our calculations, polishing the language of the manuscript and editing the figures. The authors have reviewed the AI-assisted output and take full responsibility for the content of the paper.

\appendix

\newpage
\section{Supergravity spectra}
\label{app:linear}
We review the linear harmonic decomposition used in Section \ref{sugra:section} in more detail. We use the expansion \eqref{eq:perturbations}, with the AdS radius to one, and use the five-form normalization of \eqref{sugra:iib}. Then, the linearized Einstein equation is
\begin{equation}
\begin{split}
\left(\nabla^P\nabla_M h_{NP}+\nabla^P\nabla_Nh_{MP}-\nabla^2h_{MN}-\nabla_M \nabla_N h_P^P\right)=\frac{1}{3}\Big(f_{MPQRS}\bar F_N{}^{PQRS}\\
+\bar F_{MPQRS}f_N{}^{PQRS}-4\bar F_{MAPQR}h^{AB}\bar F_{NB}{}^{PQR}\Big).
\end{split}
\label{eq:linearizedeinstein}
\end{equation}
The perturbation of the five-form self-duality equation is
\begin{equation}
f_{MNPQR}=(\bar *f_{(5)})_{MNPQR}-\frac12h_A^A\bar F_{MNPQR}+5h_{[M}^A\bar F_{|A|NPQR]}.
\label{eq:linearizedduality}
\end{equation}
Here $\bar{*}$ denotes the 10D Hodge star evaluated with the background metric. Antisymmetrization is denoted by square brackets with vertical bars excluding contracted index from the antisymmetrization.

To separate the scalar and tensor sectors, we isolate the internal metric trace
\begin{equation}
\begin{split}
h_{ij}=h_{ij}^{\mathrm{TT}}+\frac15\bar g_{ij}\pi.
\end{split}
\end{equation}
The superscript TT means the transverse-traceless part, and we can impose the internal transverse gauge
\begin{equation}
\begin{split}
\nabla^ih_{ij}^{\mathrm{TT}}&=0,\qquad \nabla^ih_{\mu i}=0,\qquad \nabla^ia_{iMNP}=0.
\end{split}
\label{eq:linearmetricgauge}
\end{equation}
We then expand the scalar part of the four-form and the internal metric in harmonics with the normalization \eqref{eq:harmonicnormalization}
\begin{equation}
\begin{split}
a_{ijkl}&=\sum_{k,I}b_k^I\epsilon_{ijkl}{}^m\nabla_mY_k^I,\qquad \pi=\sum_{k,I}\pi_k^IY_k^I,\\
h'_{\mu\nu}&=\sum_{k,I}{h'}_{\mu\nu,k}^{I}Y_k^I,\qquad h_{ij}^{\mathrm{TT}}=\sum_{k,\mathcal I}\widehat\phi_k^{\mathcal I}Y_{ij,k}^{\mathcal I}.
\end{split}
\label{eq:linearharmonicexpansion}
\end{equation}
Here $Y_{ij,k}^{\mathcal{I}}$ is the tensor spherical harmonic functions and satisfies
\begin{equation}
\nabla_{S^5}^2Y_{ij,k}^{\mathcal I}=-[k(k+4)-2]Y_{ij,k}^{\mathcal I}.
\end{equation}
For each nonconstant scalar harmonic, five-form self-duality fixes the purely AdS four-form to be the AdS Hodge dual of the differential of a scalar field denoted by $b_k^I(x)$. Projecting the internal traceless and mixed Einstein equations then gives
\begin{equation}
\begin{split}
0=\left(\frac12h'{}^\mu_{\mu,k}-\frac8{15}\pi_k\right)\left(\nabla_i\nabla_j+\frac{k(k+4)}5\bar g_{ij}\right)Y_k,
\end{split}
\label{eq:linearconstraints1}
\end{equation}
For $k\geq2$, the constraint \eqref{eq:linearconstraints1} fixes the AdS trace. Substituting it into the internal trace and five-form equations gives
\begin{equation}
\begin{split}
[\Box-k(k+4)-32]\pi_k=80k(k+4)b_k,\qquad[\Box-k(k+4)]b_k=\frac45\pi_k.
\end{split}
\label{eq:coupledlinearscalars}
\end{equation}
We diagonalize this coupled system with the scalar combinations
\begin{equation}
\widehat{s}_k=\frac{\pi_k-10(k+4)b_k}{20(k+2)},\qquad \widehat{t}_k=\frac{\pi_k+10kb_k}{20(k+2)},
\label{eq:linearscalareigenfields}
\end{equation}
which yield the masses for scalar $s_k$ and $t_k$ in Table~\ref{sugra:table}. The internal transverse-traceless projection determines the remaining mass spectra needed in this paper
\begin{equation}
(\Box-k(k+4))\phi_k=0,
\label{eq:linearscalartensoridentity}
\end{equation}
\begin{equation}
[\Box+2-k(k+4)]\varphi_{\mu\nu,k}=0.
\label{eq:lineartensormode}
\end{equation}

\section{Field redefinitions for scalars}
\label{app:scalarred}
In this appendix, we collect the quadratic supergravity sources and scalar field redefinitions required for the localized contact interactions. The conventions and field redefinition mechanism are given in Section \ref{sugra:scalars}.

Let us first focus on the perturbations of $s_k$ and $t_k$. Using the perturbative expansion in \eqref{eq:perturbations}, we extract the quadratic contributions from the terms of order $\epsilon^2$. These contributions appear as source terms in the equations of motion. For $s_k$, the sourced equations of motion can then be written as
\begin{equation}\label{eq:sperturbation2nd}
(\Box-m_1^2)\widehat s_1=\frac{(k_1+4)(k_1+5)Q_{1,1}+Q_{2,1}+(k_1+4)(\nabla_\mu Q_{3,1}^\mu+Q_{4,1})}{2(k_1+2)}.
\end{equation}
Here, $Q_{1,1}$ and $Q_{2,1}$ are the traceless-scalar and trace projections of the internal Einstein equation, while $Q^\mu_{3,1}$ and $Q_{4,1}$ come from the mixed and purely AdS components of five-form self-duality 
\begin{equation}
\begin{split}
Q_{1,1}={}&\sum_{2,3}\frac{1}{f_1(f_1-5)z_{k_1}}\Bigg[2c_{123}\nabla s_2\cdot\nabla s_3+\frac{c_{123}+d_{213}+d_{312}}{(k_2+1)(k_3+1)}\\
&\times\Bigl[k_2k_3(k_2k_3+2k_2+2k_3-2)s_2s_3+\nabla_\mu\nabla_\nu s_2\nabla^\mu\nabla^\nu s_3\Bigr]\Bigg],
\end{split}
\end{equation}
\begin{equation}
\begin{split}
Q_{2,1}={}&\sum_{2,3}\frac{-2}{5z_{k_1}(k_2+1)(k_3+1)}\Bigg[k_2k_3\Bigl[-b_{123}\bigl(1+9k_3+k_2(9+7k_3)\bigr)\\
&+a_{123}\bigl[k_2^3(4+7k_3)+k_2^2(-8-13k_3+10k_3^2)+4(-20-29k_3-2k_3^2+k_3^3)\\
&+k_2(-116-118k_3-13k_3^2+7k_3^3)\bigr]\Bigr]s_2s_3\\
&+(k_2+1)(k_3+1)\bigl[6b_{123}+5a_{123}\bigl(4k_2+4k_3-k_2k_3\bigr)\bigr]\nabla s_2\cdot\nabla s_3\\
&+\bigl[-2b_{123}+3a_{123}(k_2-k_2^2+k_3-k_3^2)\bigr]\nabla_\mu\nabla_\nu s_2\nabla^\mu\nabla^\nu s_3\Bigg],
\end{split}
\end{equation}
\begin{equation}
\begin{split}
Q_{3,1}^\mu={}&\sum_{2,3}\frac{2}{f_1z_{k_1}(k_2+1)(k_3+1)}\Big[b_{312}(k_2+1)\bigl[\nabla^\mu\nabla^\nu s_3\nabla_\nu s_2+k_3(k_3+2)s_3\nabla^\mu s_2\bigr]\\
&+b_{213}(k_3+1)\bigl[\nabla^\mu\nabla^\nu s_2\nabla_\nu s_3+k_2(k_2+2)s_2\nabla^\mu s_3\bigr]\Big],
\end{split}
\end{equation}
\begin{equation}
\begin{split}
Q_{4,1}={}&-\sum_{2,3}\frac{4a_{123}}{z_{k_1}(k_2+1)(k_3+1)}\Big[k_2k_3\bigl[(k_2+k_3)(k_2+1)(k_3+1)-k_2k_3-4\bigr]s_2s_3\\
&+\nabla_\mu\nabla_\nu s_2\nabla^\mu\nabla^\nu s_3\Big].
\end{split}
\end{equation}
We denote the scalar harmonic eigenvalues by $f_a=k_a(k_a+4)$ and suppress the degeneracy labels. The indices $1,2,3$ distinguish the modes. The relevant harmonic overlap integrals are
\begin{equation}
\begin{split}
a_{123}&=\int d\mu_5\,Y_1Y_2Y_3,\qquad b_{123}=\int d\mu_5\,Y_1\nabla_iY_2\nabla^iY_3,\\
c_{123}&=\int d\mu_5\left(\nabla_i\nabla_jY_1+\frac{f_1}{5}\bar g_{ij}Y_1\right)\nabla^iY_2\nabla^jY_3,\\
d_{123}&=\int d\mu_5\,Y_1\left(\nabla_i\nabla_jY_2+\frac{f_2}{5}\bar g_{ij}Y_2\right)\left(\nabla^i\nabla^jY_3+\frac{f_3}{5}\bar g^{ij}Y_3\right).
\end{split}
\end{equation}

To eliminate the higher-derivative terms in $Q_{1,1},\ldots,Q_{4,1}$, we use the scalar field redefinition
\begin{equation}
\widehat s_1=s_1+\sum_{2,3}\mathcal A^s_{123}\;,\qquad \mathcal A^s_{123}=J^s_{123}s_2s_3+L^s_{123}\nabla s_2\cdot\nabla s_3\;.
\end{equation}
Substitution into \eqref{eq:sperturbation2nd} separates the source into its zero-, two- and four-derivative structures. Their coefficients are
\begin{equation}
\begin{split}
D_{123}={}&-\frac{2^{k_1-3}(k_1+1)k_2k_3a_{123}}{5(k_1-1)k_1}\Big[7k_1^4-3(k_2-k_3)^2(k_2+k_3)(4+k_2+k_3)\\
&+k_1^3(46+28k_2+28k_3)+4k_1^2\bigl[-66+3k_2^2+k_3(-19+3k_3)+k_2(-19+10k_3)\bigr]\\
&+2k_1\bigl[128-6k_2^3+3k_2^2(-9+2k_3)+k_2(36+28k_3+6k_3^2)\\
&-3k_3\bigl(-12+k_3(9+2k_3)\bigr)\bigr]\Big]-\frac{1}{5}F_{123}m_2^2m_3^2,
\end{split}
\end{equation}
\begin{equation}
\begin{split}
E_{123}={}&-\frac{2^{k_1-3}(k_1+1)a_{123}}{(k_1-1)k_1}\Big[-3k_1^4-4k_1^3(-1+k_2+k_3)\\
&\qquad+(k_2-k_3)^2(k_2+k_3)(4+k_2+k_3)\\
&\qquad+2k_1^2\bigl[24+k_2^2-2k_2(-3+k_3)+k_3(6+k_3)\bigr]\\
&\qquad+4k_1\bigl[-16+k_2^3-k_2^2(-3+k_3)+(k_3-1)k_3(k_3+4)-k_2\bigl(4+k_3(7+k_3)\bigr)\bigr]\Big],
\end{split}
\end{equation}
\begin{equation}
\begin{split}
F_{123}={}&\frac{2^{k_1-4}(k_1+1)(k_1+k_2+k_3+4)a_{123}}{(k_1-1)k_1(k_2+1)(k_3+1)}\Big[k_1^3+(k_2-k_3)^2(k_2+k_3)\\
&\qquad+k_1^2(4+3k_2+3k_3)+k_1\bigl[-8+3k_2^2+k_3(-4+3k_3)-2k_2(2+3k_3)\bigr]\Big].
\end{split}
\end{equation}
Here we have introduced $m_a^2=k_a(k_a-4)$. The coefficients in the ansatz can then be determined by matching the terms without higher derivatives after the substitution
\begin{equation}
L_{123}^s=\frac{1}{2}F_{123},\qquad J^s_{123}=\frac12E_{123}+\frac{8-m_2^2-m_3^2+m_1^2}{4}F_{123}.
\end{equation}
The remaining cubic interaction then takes the canonical form without higher derivative terms \cite{Lee:1998bxa}
\begin{equation}
\begin{split}
\lambda_{123}={}&\frac{2^{k_1-6}(k_1+1)a_{123}}{(k_1-1)k_1(k_2+1)(k_3+1)}(k_1-k_2-k_3)(k_1+k_2-k_3)(k_1-k_2+k_3)\\
&\times(k_1+k_2+k_3-4)(k_1+k_2+k_3-2)(k_1+k_2+k_3)\\
&\times(k_1+k_2+k_3+2)(k_1+k_2+k_3+4).
\end{split}
\end{equation}

The same procedure applies to the other scalar towers. Here we quote the coefficients needed here for the $t_k$ tower
\begin{equation}\label{eq:fieldredefJt}
\begin{split}
J^t_{123}={}&\frac{2^{k_1-7}(k_1+1)a_{123}}{(k_1+4)(k_1+5)(k_2+1)(k_3+1)}\Big[2k_1^6+8(5-k_2-k_3)k_1^5+k_1^4\bigl[312-136(k_2+k_3)\\
&+11(k_2+k_3)^2-(k_2-k_3)^2\bigr]+k_1^3\bigl[1184-880(k_2+k_3)+160(k_2+k_3)^2\\
&-4(k_2+k_3)^3+4(k_2+k_3-4)(k_2-k_3)^2\bigr]+k_1^2\bigl[2176-2656(k_2+k_3)\\
&+840(k_2+k_3)^2-60(k_2+k_3)^3-3(k_2+k_3)^4\\
&+\bigl(-72+44(k_2+k_3)-8(k_2+k_3)^2\bigr)(k_2-k_3)^2+(k_2-k_3)^4\bigr]\\
&+k_1\bigl[1536-3584(k_2+k_3)+1816(k_2+k_3)^2-320(k_2+k_3)^3-30(k_2+k_3)^4\\
&+\bigl(-56+176(k_2+k_3)-20(k_2+k_3)^2+12(k_2+k_3)^3\bigr)(k_2-k_3)^2\\
&-\bigl(6+4(k_2+k_3)\bigr)(k_2-k_3)^4\bigr]-1536(k_2+k_3)+1248(k_2+k_3)^2-576(k_2+k_3)^3\\
&-72(k_2+k_3)^4+\bigl[160+288(k_2+k_3)+56(k_2+k_3)^2+36(k_2+k_3)^3\\
&-3(k_2+k_3)^4\bigr](k_2-k_3)^2+\bigl[(k_2+k_3)^2-12(k_2+k_3)-40\bigr](k_2-k_3)^4\Big].
\end{split}
\end{equation}
\begin{equation}\label{eq:fieldredefLt}
\begin{split}
L^t_{123}={}&\frac{2^{k_1-5}(k_1+1)(k_1-k_2-k_3)a_{123}}{(k_1+4)(k_1+5)(k_2+1)(k_3+1)}\Big[k_1^3+\bigl[8-3(k_2+k_3)\bigr]k_1^2+\bigl[8-28(k_2+k_3)\\
&+3(k_2-k_3)^2\bigr]k_1+(12-k_2-k_3)(k_2-k_3)^2-64(k_2+k_3)-32\Big].
\end{split}
\end{equation}
The $t_0$ coefficients obtained from the low harmonic equation agree with the $k_1=0$ specialization of \eqref{eq:fieldredefJt} and \eqref{eq:fieldredefLt}. Similarly, the field redefinition for $\phi_k$ is given by
\begin{equation}
\begin{split}
J^\phi_{123}&=\frac{2[-4+k_2^2+4k_2k_3+k_3^2-k_1(k_1+4)]}{(k_2+1)(k_3+1)}\int d\mu_5\,Y_{ij,1}\nabla^iY_2\nabla^jY_3,\\
L^\phi_{123}&=-\frac4{(k_2+1)(k_3+1)}\int d\mu_5\,Y_{ij,1}\nabla^iY_2\nabla^jY_3.
\end{split}
\label{eq:internaltensormap}
\end{equation}

\section{Off-shell spin-two reduction}
\label{app:spin2reduction}
We determine the spin-two field redefinition using the method of Appendix~\ref{app:scalarred}. We label the spin-two mode by $1$ and the two scalar modes that source it by $2,3$. The harmonic degeneracy labels are suppressed. Expanding \eqref{sugra:iib} to quadratic order, subtracting the AdS and sphere Einstein traces, and using the scalar five-form constraints gives
\begin{equation}
\begin{split}
\mathcal E_{\mu\nu}(\widehat\varphi_1)={}&-\frac1{z_{k_1}}\int d\mu_5\,Y_1\Bigl[(R_{\mu\nu}-T_{\mu\nu})|_{\epsilon^2}-\frac12\bar g_{\mu\nu}\bar g^{\rho\sigma}(R_{\rho\sigma}-T_{\rho\sigma})|_{\epsilon^2}\\
&\hspace{36mm}-\frac12\bar g_{\mu\nu}\bar g^{ij}(R_{ij}-T_{ij})|_{\epsilon^2}\Bigr]+12\bar g_{\mu\nu}Q_{4,1}.
\end{split}
\label{eq:spinprojection}
\end{equation}
Here $Q_{4,1}$ is the five-form self-duality source defined in Appendix~\ref{app:scalarred}. The off-shell operator retains the trace and longitudinal components of an internal spin-two propagator \cite{DHoker:1999bve,Costa:2014kfa}
\begin{equation}
\begin{split}
\mathcal E_{\mu\nu}(\varphi_1)={}&-\frac12\Box\varphi_{\mu\nu,1}+\nabla_{(\mu}\nabla^\rho\varphi_{\nu)\rho,1}+\left(\frac{f_1}{2}-1\right)\varphi_{\mu\nu,1}-\frac12\nabla_\mu\nabla_\nu\varphi^\rho_{\rho,1}\\
&-\frac12\bar g_{\mu\nu}\nabla^\rho\nabla^\sigma\varphi_{\rho\sigma,1}+\frac12\bar g_{\mu\nu}\Box\varphi^\rho_{\rho,1}-\left(\frac{f_1}{2}+1\right)\bar g_{\mu\nu}\varphi^\rho_{\rho,1}.
\end{split}
\label{eq:spinoperator}
\end{equation}
After imposing the transverse and traceless conditions, \eqref{eq:spinoperator} reduces to its on-shell form, yielding the mass spectrum of the tensor field given in \eqref{eq:lineartensormode}.

The right hand of \eqref{eq:spinprojection} contains the quadratic contributions to the spin-two equation of motion. The calculation is lengthy, and we only present the final expression
\begin{equation}
\begin{split}
\mathcal E_{\mu\nu}(\widehat\varphi_1)=&-\sum_{2,3}\frac{a_{123}}{z_{k_1}(k_2+1)(k_3+1)}\Bigg[\alpha_{123}\nabla_{(\mu}s_2\nabla_{\nu)}s_3+\beta_{123}\nabla_\rho\nabla_{(\mu}s_2\nabla^\rho\nabla_{\nu)}s_3\\
&+4\nabla_\rho\nabla^\lambda\nabla_{(\mu}s_2\nabla^\rho\nabla_\lambda\nabla_{\nu)}s_3+\mu_{123}\nabla_{(\mu}\!\left(s_3\nabla_{\nu)}s_2\right)+\mu_{132}\nabla_{(\mu}\!\left(s_2\nabla_{\nu)}s_3\right)\\
&+4(4k_2+3)\nabla_{(\mu}\!\left(\nabla^\rho s_2\nabla_{\nu)}\nabla_\rho s_3\right)+4(4k_3+3)\nabla_{(\mu}\!\left(\nabla^\rho s_3\nabla_{\nu)}\nabla_\rho s_2\right)\\
&+\bar g_{\mu\nu}\Bigl[T^1_{123}s_2s_3+T^2_{123}\nabla s_2\cdot\nabla s_3+T^3_{123}\nabla_\rho\nabla_\lambda s_2\nabla^\rho\nabla^\lambda s_3\\
&-2\nabla_\rho\nabla_\lambda\nabla_\tau s_2\nabla^\rho\nabla^\lambda\nabla^\tau s_3\Bigr]\Bigg].
\end{split}
\end{equation}
The source terms in the equations of motion for the tensor fields can likewise be classified according to their tensor structures, although the decomposition is more involved than in the scalar case. The coefficients of the resulting terms are given by
\begin{equation}
\begin{split}
\alpha_{123}=&2\Big[-f_1\bigl[(k_2-k_3)^2-3(k_2+k_3)-2\bigr]+k_2^4+k_3^4-2k_2k_3(k_2^2+k_3^2)\\
&-7(k_2^3+k_3^3)-11k_2k_3(k_2+k_3)-14(k_2^2+k_3^2)-4k_2k_3+14(k_2+k_3)+6\Big],
\end{split}
\end{equation}
\begin{equation}
\beta_{123}=-4\bigl[f_1-2k_2^2-2k_3^2+k_2+k_3+10\bigr],
\end{equation}
\begin{equation}
\mu_{123}=-2k_3\bigl[(1-k_3)f_1-(k_3+5)k_2^2+6k_2k_3+k_3^3-5k_3^2+16\bigr].
\end{equation}
\begin{equation}
\begin{split}
T^1_{123}=k_2k_3\Big[&\bigl[3k_2k_3+8(k_2+k_3)-6\bigr]f_1-160-48(k_2+k_3)+30(k_2^2+k_3^2)\\
&-8(k_2^3+k_3^3)-100k_2k_3+14k_2k_3(k_2+k_3)-k_2k_3(k_2^2+k_3^2)+4k_2^2k_3^2\Big],
\end{split}
\end{equation}
\begin{equation}
\begin{split}
T^2_{123}={}&-2f_1(k_2+1)(k_3+1)+2\Bigl[42+54(k_2+k_3)+11(k_2^2+k_3^2)+k_2^3+k_3^3\\
&+62k_2k_3+5k_2k_3(k_2+k_3)+k_2k_3(k_2^2+k_3^2)-5k_2^2k_3^2\Bigr],\\
\end{split}
\end{equation}
\begin{equation}
T^3_{123}={}3f_1-5(k_2^2+k_3^2)-10(k_2+k_3)+8.
\end{equation}

Using these source terms, we can now formulate the ansatz \eqref{sugra:spinmap1} for the field redefinition that eliminates the higher-derivative terms in the couplings between $s_k$ and the tensor fields $\varphi_{\mu\nu,k}$. The unknown coefficients in the ansatz \eqref{sugra:spinmap1} can be determined as in the scalar case. Since the expressions for general $k_2$ and $k_3$ are quite cumbersome, we present only the case $k_2=k_3=2$ below. Taking the trace of the redefined tensor equation and using the longitudinal constraint together with the free scalar equations eliminates all derivatives of the trace composite, leaving an algebraic equation that determines $\eta_1$ as
\begin{equation}
\begin{split}
\eta_1=\sum_{2,3}\frac{a_{123}}{z_{k_1}}\Biggl[\frac{f_1^3-36f_1^2+448f_1-1536}{72(f_1+3)}s_2s_3+\frac{f_1^2-8f_1+240}{36(f_1+3)}\nabla s_2\cdot\nabla s_3\Biggr].
\end{split}
\label{eq:spintracecomposite}
\end{equation}
Substituting the solution for $\eta_1$ into the longitudinal constraint determines $\xi_{\mu,1}$
\begin{equation}
\xi_{\mu,1}=\sum_{2,3}\frac{a_{123}}{z_{k_1}}\nabla_\mu\left[\frac{3f_1^2-236f_1+2496}{144(f_1+3)}s_2s_3+\frac{f_1-88}{24(f_1+3)}\nabla s_2\cdot\nabla s_3\right].
\label{eq:spinmassivecomposite}
\end{equation}
We should note that at $k_1=0$, the trace equation still determines $\eta_1$ but the constraint leaves $\xi_{\mu,1}$ undetermined which corresponds to a gauge choice for the massless graviton. We have the the degrees of freedom for gauge choice leaving a gauge independent contact diagram. 

The remaining coefficients $J_{123}^{\varphi}$ and $L_{123}^{\varphi}$ can then be determined by requiring the four- and six-derivative terms to vanish, respectively
\begin{equation}
\begin{split}
L^\varphi_{123}=\frac{4a_{123}}{9z_{k_1}}, \qquad
J^\varphi_{123}=\frac{2(30-f_1)a_{123}}{9z_{k_1}}.
\end{split}
\label{eq:spinmapcoefficients}
\end{equation}

\section{Details of the Wilson-line calculation}
\label{appWL:details}
We provide the omitted details of the Wilson-line correlator calculation in Section~\ref{wl:section}, concentrating on the $k_1=k_2=2$ case. The contact terms arise from both the nonlinear KK redefinitions and
from the collapse of worldsheet propagators. For the latter, we need the mixed vertex. It contains the displacement of the evaluation point and the variation of  the tangent vectors in the pullback metric
\begin{equation}
\begin{split}
\mathcal V_{\mathrm{F1}}^{(1,1)}={}&\frac12\sum_{k,I}\frac4{k+1}y^i\partial_iY_k^I(\theta)P_\parallel^{\alpha\beta}\left(\nabla_\alpha\nabla_\beta-\frac{k(k-1)}2\bar g_{\alpha\beta}\right)s_k^I\\
&+\frac12\hat z_0\zeta^rP_\parallel^{\alpha\beta}\partial_rh_{\alpha\beta}+\bar\gamma^{\alpha\beta}h_{\alpha r}\partial_\beta(\hat z_0\zeta^r).
\end{split}
\label{eq:wlmixedpullback}
\end{equation}
The last term arises from fluctuations of the tangent vector. Including this contribution, substituting \eqref{sugra:metric}, and integrating by parts yields \eqref{wl:mixed} for $k=2$. The kinetic operators isolated in \eqref{worldsheetibp} then act on the Green functions in \eqref{wl:Green}, producing the local contributions from worldsheet exchange. Their R-symmetry factor is
\begin{equation}
\frac{\nabla_{S^5}(u_1\cdot\theta)^2\cdot\nabla_{S^5}(u_2\cdot\theta)^2}{(u_1\cdot\theta)^2(u_2\cdot\theta)^2}=4(\sigma-1).
\label{eq:wlspherecontraction}
\end{equation}
The resulting local worldsheet-exchange contributions for two level-two external scalars are as follows. The $\zeta$ exchange gives
\begin{equation}\label{eq:wlzetacontact}
\begin{split}
\left.F_{22,\mathrm{contact}}\right|_{\zeta\ \mathrm{exchange}}&=-\frac9\pi\left[-\frac{112}9\chi C_{33}+32\chi(\xi+\chi)C_{44}\right].
\end{split}
\end{equation}
Similarly, for $y$ exchange, we obtain
\begin{equation}\label{eq:wlycontact}
\begin{split}
\left.F_{22,\mathrm{contact}}\right|_{y\ \mathrm{exchange}}&=-\frac9\pi\,4(\sigma-1)\left[-\frac{16}9C_{22}+\frac{16}9(\xi+\chi)C_{33}\right].
\end{split}
\end{equation}
Using the normalization \eqref{wl:normalization} and the Euclidean sign convention of \cite{Gimenez-Grau:2023fcy}, we obtain the overall factor
\begin{equation}
-2\frac{\sqrt\lambda}{2\pi}\left(\frac3{2\sqrt2N}\right)^2\left(\frac{\sqrt\lambda}{8N^2}\right)^{-1}=-\frac9\pi.
\end{equation}

We next consider the contact diagrams induced by field redefinitions. Substituting the scalar and spin-two field redefinitions into the metric \eqref{sugra:metric} and including the Weyl subtraction, gives
\begin{equation}
\begin{split}
&\int_{AdS_2}\sqrt{\bar\gamma}\,P_\parallel^{\mu\nu}\left.h_{\mu\nu}\right|_{s_2s_2}=\sum_{\substack{k=0,2,4\\I,J,K}}\frac{a_{kK,2I,2J}}{z_k}Y_k^K(\theta)\int_{AdS_2}\sqrt{\bar\gamma}\,\Biggl[\frac{60-2k(k+4)}9P_\parallel^{\mu\nu}\nabla_\mu s_2^I\nabla_\nu s_2^J\\
&\qquad+\frac49P_\parallel^{\mu\nu}\nabla_\rho\nabla_\mu s_2^I\nabla^\rho\nabla_\nu s_2^J+\frac{304-12k(k+4)}9s_2^Is_2^J-\frac89\nabla s_2^I\cdot\nabla s_2^J\Biggr].
\end{split}
\label{eq:wlcommonmetricterms}
\end{equation}
The R-symmetry polynomials arising from the field redefinitions are less straightforward to determine than those associated with propagator collapsing. In particular, one must decompose the product of the two spherical harmonics. The normalized projections evaluated on the string worldsheet are given by
\begin{equation}
(u_1\cdot\theta)^2(u_2\cdot\theta)^2\mathcal{R}_k=\sum_I\frac{Y_k^I(\theta)}{z_k}\int d\mu_5\,Y_k^I(n)(u_1\cdot n)^2(u_2\cdot n)^2,
\end{equation}
where we have restored the factor $a_{kK,2I,2J}$ in \eqref{eq:wlcommonmetricterms} in the integral representation. We can  evaluate all possible harmonic projections and get
\begin{equation}
\label{eq:wlharmonicprojections}
\begin{split}
\mathcal{R}_0^{(2,2)}=\frac{\sigma^2}{24},\qquad\mathcal{R}_2^{(2,2)}=\frac{2}{5}\sigma-\frac{1}{15}\sigma^2,\qquad \mathcal{R}_4^{(2,2)}=1-\frac25\sigma+\frac1{40}\sigma^2.
\end{split}
\end{equation}

From the field-redefinition-induced contact interactions in \eqref{eq:wlcommonmetricterms}, we see that the relevant Witten diagrams involve AdS derivative. The contact diagram in which two derivatives are contracted with the metric tangent to the F1 worldsheet can be evaluated directly
\begin{equation}\label{eq:wlfr2der}
\begin{split}
|\mathbf{x}_{1\perp}|^2|\mathbf{x}_{2\perp}|^2\int_{AdS_2}\sqrt{\bar\gamma}\,P_\parallel^{\mu\nu}\nabla_\mu K_2(z,x_1)\nabla_\nu K_2(z,x_2)&=4C_{22}-8(\xi+\chi)C_{33}.
\end{split}
\end{equation}
It remains to evaluate the contact Witten diagrams induced by the field redefinitions that involve four derivatives. These diagrams can be computed using the following identity
\begin{equation}
\nabla_\mu\nabla_\nu K_\Delta=\Delta(\Delta+1)\,K_1^{\Delta-2}\,\nabla_\mu K_1\,\nabla_\nu K_1-\Delta\,\bar g_{\mu\nu}K_\Delta .
\end{equation}
We can then explicitly evaluate the derivative contractions in the four derivative contact diagram for $k_1=k_2=2$
\begin{equation}
\begin{split}
P_\parallel^{\mu\nu}&\nabla_\rho\nabla_\mu K_2(z,x_1)\nabla^\rho\nabla_\nu K_2(z,x_2)=36P_\parallel^{\mu\nu}\nabla_\mu K_1(z,x_1)\nabla_\nu K_1(z,x_2)\\
&\times\nabla_\rho K_1(z,x_1)\nabla^\rho K_1(z,x_2)-12K_2(z,x_2)\,P_\parallel^{\mu\nu}\nabla_\mu K_1(z,x_1)\nabla_\nu K_1(z,x_1)\\
&-12K_2(z,x_1)\,P_\parallel^{\mu\nu}\nabla_\mu K_1(z,x_2)\nabla_\nu K_1(z,x_2)+8K_2(z,x_1)K_2(z,x_2).
\end{split}
\end{equation}
The remaining tangential contraction can be evaluated using
\begin{equation}
\begin{split}
P_\parallel^{\mu\nu}\nabla_\mu K_1(z,x_1)\nabla_\nu K_1(z,x_2)=&K_1(z,x_1)K_1(z,x_2)\\
&-2(\xi+\chi)\,|\mathbf x_{1\perp}||\mathbf x_{2\perp}|\,K_2(z,x_1)K_2(z,x_2).
\end{split}
\end{equation}
\begin{equation}
P_\parallel^{\mu\nu}\nabla_\mu K_1(z,x_i)\nabla_\nu K_1(z,x_i)=K_2(z,x_i)-4|\mathbf x_{i\perp}|^2K_4(z,x_i).
\end{equation}
We can then write the four-derivative contact diagram as
\begin{equation}
\begin{split}
|\mathbf{x}_{1\perp}|^2|\mathbf{x}_{2\perp}|^2\int_{AdS_2}\sqrt{\bar\gamma}\,P_\parallel^{\mu\nu}\nabla_\rho\nabla_\mu K_2(z,x_1)\nabla^\rho\nabla_\nu K_2(z,x_2)&=20C_{22}+48(C_{24}+C_{42})\\
&\hspace{-27mm}-72(2\xi+\chi)C_{33}+144\xi(\xi+\chi)C_{44}.
\end{split}
\label{eq:wlderivativereduction}
\end{equation}
Combining the R-symmetry polynomials in \eqref{eq:wlharmonicprojections} with the Witten diagrams in \eqref{eq:wlfr2der} and \eqref{eq:wlderivativereduction}, we obtain the contributions of the field redefinitions to the correlation functions
\begin{equation}
\begin{split}
\left.F_{22,\mathrm{contact}}\right|_{\mathrm{field\ redefinitions}}=&-\frac9\pi\Biggl[\frac{80\sigma-24}9C_{22}+\frac{32}3(C_{24}+C_{42})\\
&-\frac{240\xi+128\chi+64\sigma(\xi+\chi)}9C_{33}+32\xi(\xi+\chi)C_{44}\Biggr].
\end{split}
\label{eq:wlfieldredefinitioncontact}
\end{equation}

The contact diagram arising from the determinant term in \eqref{wl:determinant} is straightforward to evaluate, so we present only the final expression
\begin{equation}\label{eq:wldetercontact}
\left.F_{22,\mathrm{contact}}\right|_{\mathrm{determinant}}=-\frac9\pi\left[-\frac83C_{22}+\frac{32}3(C_{24}+C_{42})+\frac{32}9(\xi+\chi)C_{33}-128C_{44}\right].
\end{equation}
Combining \eqref{eq:wlzetacontact}, \eqref{eq:wlycontact}, \eqref{eq:wlfieldredefinitioncontact}, and \eqref{eq:wldetercontact} yields the decomposition in \eqref{wl:contact-split} and the total contact contribution in \eqref{wl:contact-total}.

We next consider the bulk channel exchange diagrams, beginning with scalar exchange. With the standard AdS boundary conditions, the Green function reads
\begin{equation}\label{eq:btobpropagator}
\begin{split}
G_\Delta(z,w)=&\frac{\Gamma(\Delta)}{2\pi^2\Gamma(\Delta-1)}\left(\frac{z_0w_0}{(z_0-w_0)^2+|\mathbf z-\mathbf w|^2}\right)^{\!\Delta}\\
&\times{}_2F_1\!\left(\Delta,\Delta-\frac32;\,2\Delta-3;\,-\frac{4z_0w_0}{(z_0-w_0)^2+|\mathbf z-\mathbf w|^2}\right).
\end{split}
\end{equation}
The bulk channel scalar exchange Witten diagram for $k_1=k_2=2$ is then given by
\begin{equation}
\frac{\mathcal E_{22}^{2,0}}{|\mathbf{x}_{1\perp}|^2|\mathbf{x}_{2\perp}|^2}=\int_{AdS_2}\sqrt{\bar\gamma}\int_{AdS_5}\sqrt{\bar g}\,K_2(z,x_1)K_2(z,x_2)G_2(z,\hat z).
\end{equation}
The bulk equation of motion reduces $\mathcal E_{22}^{2,0}$ to contact diagrams \cite{DHoker:1999mqo}
\begin{equation}
\mathcal E_{22}^{2,0}=\frac{C_{11}}{4\xi}.
\end{equation}

For massless spin two exchange, the cubic vertex \eqref{sugra:spinvertex} supplies the conserved source
\begin{equation}
\begin{split}
T_{\mu\nu}&=\nabla_{(\mu}K_2(z,x_1)\nabla_{\nu)}K_2(z,x_2)-\frac12\bar g_{\mu\nu}\left[\nabla K_2(z,x_1)\cdot\nabla K_2(z,x_2)-4K_2(z,x_1)K_2(z,x_2)\right].
\end{split}
\label{eq:wlspinsource}
\end{equation}
We emphasize that, in the exchange diagram calculation, the spin-two fields are off shell, and the complete propagators must therefore be used. Contracting the propagator with the source and the linearized string vertex yields the spin two exchange diagram
\begin{equation}
\frac{\mathcal E_{22}^{4,2}}{|\mathbf{x}_{1\perp}|^2|\mathbf{x}_{2\perp}|^2}=\frac12\int_{AdS_2}\sqrt{\bar\gamma}\int_{AdS_5}\sqrt{\bar g}\,P_\parallel^{\mu\nu}G_{\mu\nu;\rho\sigma}^{4,2}(\hat z,z)T^{\rho\sigma}(z).
\label{eq:wlspinexchange}
\end{equation}
The spin two exchange diagram does not require explicit evaluation of the bulk integral. Instead, we can use the equations of motion to reduce it to a finite sum of contact diagrams \cite{Arutyunov:2002fh,Berdichevsky:2007xd,Gimenez-Grau:2023fcy}
\begin{equation}
\mathcal E_{22}^{4,2}=-\frac{2}{3\xi}\left(\frac\chi2C_{22}-C_{13}\right)-\frac{1}{3}C_{22}.
\label{eq:wlspinconversion}
\end{equation}
The first term agrees with the exchange contribution defined using transverse contractions in \cite{Gimenez-Grau:2023fcy}. Rewriting it in terms of tangential contractions introduces an additional trace contribution, since the tangential projector satisfies $P_{\parallel}^{\mu\nu}=g^{\mu\nu}-P_{\perp}^{\mu\nu}$. The resulting contact term forms part of the Regge improvement discussed in \cite{Gimenez-Grau:2023fcy}.

Two fields are exchanged in the defect channel. The first exchanged field is $y^i$ and gives
\begin{equation}
\frac{\widehat{\mathcal E}_{22}^{1,0}}{|\mathbf{x}_{1\perp}|^2|\mathbf{x}_{2\perp}|^2}=\int_{AdS_2}d^2\hat z\,d^2\hat w\,\sqrt{\bar\gamma(\hat z)\bar\gamma(\hat w)}\,K_2(\hat z,x_1)\widehat G_1(\hat z,\hat w)K_2(\hat w,x_2),
\end{equation}
The second exchanged field is $\zeta$ and gives
\begin{equation}
\frac{\widehat{\mathcal E}_{22}^{2,1}}{|\mathbf{x}_{1\perp}|^2|\mathbf{x}_{2\perp}|^2}=\int_{AdS_2}d^2\hat z\,d^2\hat w\,\sqrt{\bar\gamma(\hat z)\bar\gamma(\hat w)}\,\hat z_0\partial_rK_2(\hat z,x_1)\widehat G_2(\hat z,\hat w)\hat w_0\partial_rK_2(\hat w,x_2).
\end{equation}
Here we normalize the worldsheet propagators by
\begin{equation}
\widehat G_{\Delta}(\hat z,\hat w)=\frac{\Gamma(\Delta)}{2\sqrt{\pi}\,\Gamma\!\left(\Delta+\tfrac12\right)}\left(\frac{\hat z_0\hat w_0}{|\hat z-\hat w|^2}\right)^{\!\Delta}{}_2F_1\!\left(\Delta,\Delta;\,2\Delta;\,-\frac{4\hat z_0\hat w_0}{|\hat z-\hat w|^2}
\right).
\end{equation}
For the scalar exchange, the equations of motion can likewise be used to reduce the defect channel exchange diagram to a contact diagram \cite{Gimenez-Grau:2023fcy,Chen:2025yxg,Chen:2026ium}
\begin{equation}
\left[-\bigl((\xi+\chi)^2-4\bigr)\partial_\xi^2-2(\xi+\chi)\partial_\xi+\widehat\Delta(\widehat\Delta-1)\right]\widehat{\mathcal E}_{\Delta_1\Delta_2}^{\widehat\Delta,0}=C_{\Delta_1\Delta_2}.
\end{equation}
The solution to the differential equation is fixed by imposing regularity in the defect OPE limit. After accounting for the normalization, the solution for $k_1=k_2=2$ is
\begin{equation}
\widehat{\mathcal E}_{22}^{1,0}(\xi,\chi)=\frac{\pi}{8}\left[\log(2+\xi+\chi)-\frac{\xi+\chi}{\sqrt{(\xi+\chi)^2-4}}\,\text{arccosh}\!\left(\frac{\xi+\chi}{2}\right)\right].
\end{equation}
The defect channel vector exchange follows by acting with transverse derivatives on the scalar exchange
\begin{equation}
\begin{split}
\widehat{\mathcal E}_{22}^{2,1}(\xi,\chi)=\frac{\pi\chi}{8}\Bigg[&(\xi+\chi)\log(2+\xi+\chi)-2+\frac{\xi+\chi}{(\xi+\chi)^2-4}\\
&-\frac{(\xi+\chi)^4-6(\xi+\chi)^2+12}{\bigl[(\xi+\chi)^2-4\bigr]^{3/2}}\text{arccosh}\!\left(\frac{\xi+\chi}{2}\right)\Bigg].
\end{split}
\end{equation}

\section{Details of the O1 calculation}
\label{appO1:details}
We give the details of the O1 two-point functions for $k_1=k_2=2$
and $k_1=k_2=3$. We first evaluate the contact interactions induced
by the field redefinitions in Appendix~\ref{app:scalarred}. We then
evaluate the exchange diagrams by the method of images
\cite{Giombi:2020xah}.

We first consider the contact contribution for $k_1=k_2=2$. The induced O1 coupling in \eqref{o1:induced} receives contributions from the field redefinitions of $s_k$, $t_k$, and $\phi_k$. After collecting all contributions, we obtain \eqref{o1:quadratic}. Our task is then to calculate the $22$ contact diagram directly from \eqref{o1:quadratic}. For the one-point function, the wrapped integral appearing in the squared term therefore gives
\begin{equation}
\begin{split}
\int_{S^2}d\Omega_2\,K_2^{\mathrm{O1}}(z_\star,x)&=\frac{3}{\sqrt{2}\,N}\,K_2(z_\star,x)\int_{S^2}d\Omega_2\,(u\cdot n)^2=\frac{\sqrt{2}\,\pi}{N}\,\frac{u\cdot\bar u}{(1+x^2)^2}.
\end{split}
\end{equation}
Using the one-point function, the $22$ contact diagram can be constructed from its product with the appropriate symmetry factor due to \eqref{o1:quadratic}
\begin{equation}
\mathcal G_{22,\mathrm{contact}}=\frac{8\pi T_{\mathrm{O1}}}{N^2}.
\end{equation}

For the case $k_1=k_2=3$, the contact interactions induced by the field redefinitions can be obtained similarly. However, we do not have a compact quadratic action analogous to \eqref{o1:quadratic}. We therefore consider the field redefinitions of $s_k$, $t_k$, and $\phi_k$ separately. The contribution from the field redefinition of $s_k$ is given by
\begin{equation}
\begin{split}
\mathcal G^s_{33,\mathrm{contact}}=\frac{4\pi T_{\mathrm{O1}}}{N^2}\Biggl[&\frac{\sigma+\bar\sigma}{150} \Bigl(1538+527(\sigma^2+\bar\sigma^2)+407\sigma\bar\sigma\Bigr)\\
&-\frac{9}{350}(1-2\eta)(\sigma-\bar\sigma)\Bigl(32+103(\sigma^2+\bar\sigma^2)+207\sigma\bar\sigma\Bigr)\Biggr].
\end{split}
\end{equation}
Similarly, the contact contributions induced by the field redefinitions of $t_k$ and $\phi_k$ are
\begin{equation}
\begin{split}
\mathcal G^t_{33,\mathrm{contact}}=&\frac{4\pi T_{\mathrm{O1}}}{N^2}\Biggl[-(\sigma+\bar\sigma)\left(3+\frac{237}{50}(\sigma^2+\bar\sigma^2)+\frac{1371}{350}\sigma\bar\sigma\right)\\
&+(1-2\eta)(\sigma-\bar\sigma)\left(\frac{12}{35}+\frac{759}{350}(\sigma^2+\bar\sigma^2)+\frac{177}{50}\sigma\bar\sigma\right)\Biggr],
\end{split}
\end{equation}
\begin{equation}
\begin{split}
\mathcal G^\phi_{33,\mathrm{contact}}=&\frac{4\pi T_{\mathrm{O1}}}{N^2}\Biggl[(\sigma+\bar\sigma)\left(-\frac{244}{75}+\frac{92}{75}(\sigma^2+\bar\sigma^2)+\frac{632}{525}\sigma\bar\sigma\right)\\
&+(1-2\eta)(\sigma-\bar\sigma)\left(\frac{12}{25}(1+\sigma^2+\bar\sigma^2)+\frac{312}{175}\sigma\bar\sigma\right)\Biggr].
\end{split}
\end{equation}
Collecting all contributions, we find that the dependence on the nontrivial spacetime cross-ratios cancels exactly, yielding a compact expression
\begin{equation}
\mathcal{G}_{33,\mathrm{contact}}=\mathcal G^s_{33,\mathrm{contact}}+\mathcal G^t_{33,\mathrm{contact}}+\mathcal G^\phi_{33,\mathrm{contact}}=\frac{16\pi T_{\mathrm{O1}}}{N^2}(\sigma+\bar\sigma).
\end{equation}

The remaining contributions arise from the exchange of bulk scalar fields. We use the propagators with the normalization \eqref{eq:btobpropagator}, and define
\begin{equation}
\mathcal W_{k_1k_2;k}(\eta)=\frac{2^{k_1+k_2-2}}{K_{k_1}(z_\star,x_1)\,K_{k_2}(z_\star,x_2)}\int_{\mathrm{AdS}_5} d^5z\,\sqrt{\bar g}K_{k_1}(z,x_1)\,K_{k_2}(z,x_2)\,G_k(z,z_\star).
\label{eq:o1exchangeintegral}
\end{equation}
The bulk exchange diagrams can also be evaluated using the equations of motion together with the method of images described in \cite{Giombi:2020xah}. The image contributions can then be obtained from the direct channel contributions through
\begin{equation}
\mathcal G_{pp,\mathrm{image}}(\eta,\sigma,\bar\sigma)=\mathcal G_{pp,\mathrm{direct}}(1-\eta,\bar\sigma,\sigma).
\label{eq:o1imageexchange}
\end{equation}
It is therefore enough to evaluate the direct channel exchange diagram. The bulk equation of motion gives
\begin{equation}
\begin{split}
\Bigl[
&4\eta^2(1-\eta)\partial_\eta^2
+4\eta\bigl[(k_1+k_2-1)-(k_1+k_2+1)\eta\bigr]\partial_\eta\\
&+(k_1+k_2)(k_1+k_2-4)
-4k_1k_2\eta-k(k-4)
\Bigr]\mathcal W_{k_1k_2;k}(\eta)
=-2^{k_1+k_2-2},
\end{split}
\label{eq:o1exchangeequation}
\end{equation}
and the solution is fixed by regularity and required falloff behavior in the OPE limit. The exchange diagram for $k_1=k_2=2$ is 
\begin{equation}
\mathcal{W}_{22,2}=\frac{1}{\eta}.
\end{equation}
For $k_1=k_2=3$, two fields can be exchanged in the bulk channel, and their contributions are given by
\begin{equation}
\mathcal W_{33;2}=\frac{1+\eta}{\eta^2},\qquad \mathcal W_{33;4}=\frac1\eta\;.
\end{equation}
Accounting for the normalization factors and carefully incorporating the R-symmetry polynomials derived in \cite{Zhou:2024ekb}, we obtain the direct channel exchange diagrams
\begin{equation}
\mathcal G_{22,\mathrm{direct}}=\frac{16\pi T_{\mathrm{O1}}}{N^2}\,\sigma\bar\sigma\,\mathcal W_{22;2}.
\end{equation}
\begin{equation}
\begin{split}
\mathcal G_{33,\mathrm{direct}}&=\frac{4\pi T_{\mathrm{O1}}}{N^2}\left[6\sigma^2\bar\sigma\,\mathcal W_{33;2}+2\sigma(2-\sigma^2+4\bar\sigma^2)\mathcal W_{33;4}\right].
\end{split}
\label{eq:o1directexchanges}
\end{equation}
Combining the direct and image exchanges with the contact terms reproduces the two correlators in Section~\ref{o1:section}.

\bibliography{refs}

@article{Linardopoulos:2026mut,
    author = "Linardopoulos, Georgios and Park, Chanyong",
    title = "{Heavy holographic correlators in defect conformal field theories}",
    eprint = "2601.15736",
    archivePrefix = "arXiv",
    primaryClass = "hep-th",
    reportNumber = "APCTP Pre2025-009",
    month = "1",
    year = "2026"
}

@article{Karch:2000gx,
    author = "Karch, Andreas and Randall, Lisa",
    title = "{Open and closed string interpretation of SUSY CFT's on branes with boundaries}",
    eprint = "hep-th/0105132",
    archivePrefix = "arXiv",
    reportNumber = "MIT-CTP-3146",
    doi = "10.1088/1126-6708/2001/06/063",
    journal = "JHEP",
    volume = "06",
    pages = "063",
    year = "2001"
}

@article{Karch:2001cw,
    author = "Karch, Andreas and Randall, Lisa",
    title = "{Localized gravity in string theory}",
    eprint = "hep-th/0105108",
    archivePrefix = "arXiv",
    doi = "10.1103/PhysRevLett.87.061601",
    journal = "Phys. Rev. Lett.",
    volume = "87",
    pages = "061601",
    year = "2001"
}

@article{Bianchi:2026ukq,
    author = "Bianchi, Lorenzo and Bonomi, Davide and Ferragatta, Maddalena and Forini, Valentina",
    title = "{Surface defects in $\mathcal{N}=4$ Super Yang-Mills: exact results and holographic bootstrap}",
    eprint = "2610.10723",
    archivePrefix = "arXiv",
    primaryClass = "hep-th",
    month = "10",
    year = "2026"
}

@article{DHoker:1999bve,
    author = "D'Hoker, Eric and Freedman, Daniel Z. and Mathur, Samir D. and Matusis, Alec and Rastelli, Leonardo",
    title = "{Graviton and gauge boson propagators in AdS(d+1)}",
    eprint = "hep-th/9902042",
    archivePrefix = "arXiv",
    reportNumber = "MIT-CTP-2824, UCLA-99-TEP-1",
    doi = "10.1016/S0550-3213(99)00524-6",
    journal = "Nucl. Phys. B",
    volume = "562",
    pages = "330--352",
    year = "1999"
}

@article{Rey:1998ik,
    author = "Rey, Soo-Jong and Yee, Jung-Tay",
    title = "{Macroscopic strings as heavy quarks in large N gauge theory and anti-de Sitter supergravity}",
    eprint = "hep-th/9803001",
    archivePrefix = "arXiv",
    reportNumber = "SNUTP-98-016",
    doi = "10.1007/s100520100799",
    journal = "Eur. Phys. J. C",
    volume = "22",
    pages = "379--394",
    year = "2001"
}

@article{Maldacena:1998im,
    author = "Maldacena, Juan Martin",
    title = "{Wilson loops in large N field theories}",
    eprint = "hep-th/9803002",
    archivePrefix = "arXiv",
    reportNumber = "HUTP-98-A014",
    doi = "10.1103/PhysRevLett.80.4859",
    journal = "Phys. Rev. Lett.",
    volume = "80",
    pages = "4859--4862",
    year = "1998"
}

@article{Schwarz:1983qr,
    author = "Schwarz, John H.",
    editor = "Salam, A. and Sezgin, E.",
    title = "{Covariant Field Equations of Chiral N=2 D=10 Supergravity}",
    reportNumber = "CALT-68-1016",
    doi = "10.1016/0550-3213(83)90192-X",
    journal = "Nucl. Phys. B",
    volume = "226",
    pages = "269",
    year = "1983"
}

@article{Howe:1983sra,
    author = "Howe, Paul S. and West, Peter C.",
    title = "{The Complete N=2, D=10 Supergravity}",
    reportNumber = "Print-83-0565 (KING'S COLL)",
    doi = "10.1016/0550-3213(84)90472-3",
    journal = "Nucl. Phys. B",
    volume = "238",
    pages = "181--220",
    year = "1984"
}

@article{Arutyunov:1999fb,
    author = "Arutyunov, G. and Frolov, S.",
    title = "{Scalar quartic couplings in type IIB supergravity on AdS(5) x S**5}",
    eprint = "hep-th/9912210",
    archivePrefix = "arXiv",
    reportNumber = "LMU-TPW-99-23, UAHEP-996",
    doi = "10.1016/S0550-3213(00)00210-8",
    journal = "Nucl. Phys. B",
    volume = "579",
    pages = "117--176",
    year = "2000"
}

@article{Kim:1985ez,
	author = {Kim, H. J. and Romans, L. J. and van Nieuwenhuizen, P.},
	doi = {10.1103/PhysRevD.32.389},
	journal = {Phys. Rev. D},
	pages = {389},
	reportnumber = {ITP-SB-84-87},
	title = {{The Mass Spectrum of Chiral N=2 D=10 Supergravity on S**5}},
	volume = {32},
	year = {1985}}

@article{Chen:2023yvw,
	archiveprefix = {arXiv},
	author = {Chen, Junding and Gimenez-Grau, Aleix and Zhou, Xinan},
	doi = {10.1103/PhysRevD.109.L061903},
	eprint = {2310.19230},
	journal = {Phys. Rev. D},
	number = {6},
	pages = {L061903},
	primaryclass = {hep-th},
	title = {{Defect two-point functions in 6D (2,0) theories}},
	volume = {109},
	year = {2024}}

@article{Rastelli:2016nze,
	archiveprefix = {arXiv},
	author = {Rastelli, Leonardo and Zhou, Xinan},
	doi = {10.1103/PhysRevLett.118.091602},
	eprint = {1608.06624},
	journal = {Phys. Rev. Lett.},
	number = {9},
	pages = {091602},
	primaryclass = {hep-th},
	slaccitation = {%%CITATION = ARXIV:1608.06624;%%},
	title = {{Mellin amplitudes for $AdS_5\times S^5$}},
	volume = {118},
	year = {2017}}

@article{Rastelli:2017ecj,
	archiveprefix = {arXiv},
	author = {Rastelli, Leonardo and Zhou, Xinan},
	doi = {10.1007/JHEP10(2017)146},
	eprint = {1705.05362},
	journal = {JHEP},
	pages = {146},
	primaryclass = {hep-th},
	reportnumber = {YITP-SB-2017-18},
	slaccitation = {%%CITATION = ARXIV:1705.05362;%%},
	title = {{The Mellin Formalism for Boundary CFT$_d$}},
	volume = {10},
	year = {2017}}

@article{Rastelli:2017udc,
	archiveprefix = {arXiv},
	author = {Rastelli, Leonardo and Zhou, Xinan},
	doi = {10.1007/JHEP04(2018)014},
	eprint = {1710.05923},
	journal = {JHEP},
	pages = {014},
	primaryclass = {hep-th},
	reportnumber = {YITP-SB-2017-44},
	slaccitation = {%%CITATION = ARXIV:1710.05923;%%},
	title = {{How to Succeed at Holographic Correlators Without Really Trying}},
	volume = {04},
	year = {2018}}

@article{Lee:1998bxa,
	archiveprefix = {arXiv},
	author = {Lee, Sangmin and Minwalla, Shiraz and Rangamani, Mukund and Seiberg, Nathan},
	doi = {10.4310/ATMP.1998.v2.n4.a1},
	eprint = {hep-th/9806074},
	journal = {Adv. Theor. Math. Phys.},
	pages = {697--718},
	reportnumber = {PUPT-1796, IASSNS-HEP-98-51, PUPT-1796-},
	title = {{Three point functions of chiral operators in D = 4, N=4 SYM at large N}},
	volume = {2},
	year = {1998}}

@article{Arutyunov:1998hf,
    author = "Arutyunov, G. E. and Frolov, S. A.",
    title = "{Quadratic action for Type IIB supergravity on AdS(5) x S**5}",
    eprint = "hep-th/9811106",
    archivePrefix = "arXiv",
    reportNumber = "UAHEP-988",
    doi = "10.1088/1126-6708/1999/08/024",
    journal = "JHEP",
    volume = "08",
    pages = "024",
    year = "1999"
}

@article{Giombi:2020xah,
	archiveprefix = {arXiv},
	author = {Giombi, Simone and Khanchandani, Himanshu and Zhou, Xinan},
	doi = {10.1088/1751-8121/abcf59},
	eprint = {2009.03290},
	journal = {J. Phys. A},
	number = {2},
	pages = {024003},
	primaryclass = {hep-th},
	title = {{Aspects of CFTs on Real Projective Space}},
	volume = {54},
	year = {2021}}

@article{Freedman:1998tz,
	archiveprefix = {arXiv},
	author = {Freedman, Daniel Z. and Mathur, Samir D. and Matusis, Alec and Rastelli, Leonardo},
	doi = {10.1016/S0550-3213(99)00053-X},
	eprint = {hep-th/9804058},
	journal = {Nucl. Phys.},
	pages = {96-118},
	primaryclass = {hep-th},
	reportnumber = {MIT-CTP-2727},
	slaccitation = {%%CITATION = HEP-TH/9804058;%%},
	title = {{Correlation functions in the CFT(d) / AdS(d+1) correspondence}},
	volume = {B546},
	year = {1999}}

@article{Berdichevsky:2007xd,
    author = "Berdichevsky, Leon and Naaijkens, Pieter",
    title = "{Four-point functions of different-weight operators in the AdS/CFT correspondence}",
    eprint = "0709.1365",
    archivePrefix = "arXiv",
    primaryClass = "hep-th",
    reportNumber = "ITP-UU-07-47, SPIN-07-35",
    doi = "10.1088/1126-6708/2008/01/071",
    journal = "JHEP",
    volume = "01",
    pages = "071",
    year = "2008"
}

@article{DHoker:1999mqo,
	archiveprefix = {arXiv},
	author = {D'Hoker, Eric and Freedman, Daniel Z. and Rastelli, Leonardo},
	doi = {10.1016/S0550-3213(99)00526-X},
	eprint = {hep-th/9905049},
	journal = {Nucl. Phys.},
	pages = {395-411},
	primaryclass = {hep-th},
	reportnumber = {MIT-CTP-2857, UCLA-99-TEP-16},
	slaccitation = {%%CITATION = HEP-TH/9905049;%%},
	title = {{AdS / CFT four point functions: How to succeed at z integrals without really trying}},
	volume = {B562},
	year = {1999}}

@article{DHoker:1999pj,
	archiveprefix = {arXiv},
	author = {D'Hoker, Eric and Freedman, Daniel Z. and Mathur, Samir D. and Matusis, Alec and Rastelli, Leonardo},
	doi = {10.1016/S0550-3213(99)00525-8},
	eprint = {hep-th/9903196},
	journal = {Nucl. Phys.},
	pages = {353-394},
	primaryclass = {hep-th},
	reportnumber = {MIT-CTP-2843, UCLA-99-TEP-2},
	slaccitation = {%%CITATION = HEP-TH/9903196;%%},
	title = {{Graviton exchange and complete four point functions in the AdS / CFT correspondence}},
	volume = {B562},
	year = {1999}}

@article{Costa:2014kfa,
	archiveprefix = {arXiv},
	author = {Costa, Miguel S. and Gon{\c c}alves, Vasco and Penedones, Jo{\~a}o},
	doi = {10.1007/JHEP09(2014)064},
	eprint = {1404.5625},
	journal = {JHEP},
	pages = {064},
	primaryclass = {hep-th},
	slaccitation = {%%CITATION = ARXIV:1404.5625;%%},
	title = {{Spinning AdS Propagators}},
	volume = {09},
	year = {2014}}

@article{Arutyunov:1999en,
	archiveprefix = {arXiv},
	author = {Arutyunov, G. and Frolov, S.},
	doi = {10.1103/PhysRevD.61.064009},
	eprint = {hep-th/9907085},
	journal = {Phys. Rev. D},
	pages = {064009},
	reportnumber = {LMU-TPW-99-12, UAHEP-99-3},
	title = {{Some cubic couplings in type IIB supergravity on AdS(5) x S**5 and three point functions in SYM(4) at large N}},
	volume = {61},
	year = {2000}}

@article{Arutyunov:2000py,
	archiveprefix = {arXiv},
	author = {Arutyunov, G. and Frolov, S.},
	doi = {10.1103/PhysRevD.62.064016},
	eprint = {hep-th/0002170},
	journal = {Phys. Rev.},
	pages = {064016},
	primaryclass = {hep-th},
	reportnumber = {LMU-TPW-2000-7, UAHEP-00-2, LMU-TPW-007, UAHEP002},
	slaccitation = {%%CITATION = HEP-TH/0002170;%%},
	title = {{Four point functions of lowest weight CPOs in N=4 SYM(4) in supergravity approximation}},
	volume = {D62},
	year = {2000}}

@article{Arutyunov:2002fh,
	archiveprefix = {arXiv},
	author = {Arutyunov, G. and Dolan, F.A. and Osborn, H. and Sokatchev, E.},
	doi = {10.1016/S0550-3213(03)00448-6},
	eprint = {hep-th/0212116},
	journal = {Nucl. Phys. B},
	pages = {273--324},
	reportnumber = {AEI-2002-096, DAMTP-02-148, LAPTH-953-02},
	title = {{Correlation functions and massive Kaluza-Klein modes in the AdS / CFT correspondence}},
	volume = {665},
	year = {2003}}

@article{Bissi:2022mrs,
	archiveprefix = {arXiv},
	author = {Bissi, Agnese and Sinha, Aninda and Zhou, Xinan},
	doi = {10.1016/j.physrep.2022.09.004},
	eprint = {2202.08475},
	journal = {Phys. Rept.},
	pages = {1--89},
	primaryclass = {hep-th},
	title = {{Selected topics in analytic conformal bootstrap: A guided journey}},
	volume = {991},
	year = {2022}}

@article{Gimenez-Grau:2023fcy,
	archiveprefix = {arXiv},
	author = {Gimenez-Grau, Aleix},
	eprint = {2306.11896},
	month = {6},
	primaryclass = {hep-th},
	title = {{The Witten Diagram Bootstrap for Holographic Defects}},
	year = {2023}}

@article{Meneghelli:2022gps,
	archiveprefix = {arXiv},
	author = {Meneghelli, Carlo and Tr\'epanier, Maxime},
	doi = {10.1007/JHEP07(2023)165},
	eprint = {2212.05020},
	journal = {JHEP},
	pages = {165},
	primaryclass = {hep-th},
	title = {{Bootstrapping string dynamics in the 6d $\mathcal{N}$ = (2, 0) theories}},
	volume = {07},
	year = {2023}}

@article{Chen:2023oax,
	archiveprefix = {arXiv},
	author = {Chen, Junding and Zhou, Xinan},
	doi = {10.1007/JHEP09(2023)204},
	eprint = {2304.11799},
	journal = {JHEP},
	pages = {204},
	primaryclass = {hep-th},
	title = {{Aspects of higher-point functions in BCFT$_{d}$}},
	volume = {09},
	year = {2023}}

@article{Barrat:2021yvp,
	archiveprefix = {arXiv},
	author = {Barrat, Julien and Gimenez-Grau, Aleix and Liendo, Pedro},
	doi = {10.1007/JHEP04(2022)093},
	eprint = {2108.13432},
	journal = {JHEP},
	pages = {093},
	primaryclass = {hep-th},
	reportnumber = {HU-EP-21/29-RTG, DESY-21-133},
	title = {{Bootstrapping holographic defect correlators in $ \mathcal{N} $ = 4 super Yang-Mills}},
	volume = {04},
	year = {2022}}

@article{Zhou:2024ekb,
	archiveprefix = {arXiv},
	author = {Zhou, Xinan},
	doi = {10.1103/PhysRevLett.133.201602},
	eprint = {2408.04926},
	journal = {Phys. Rev. Lett.},
	number = {20},
	pages = {201602},
	primaryclass = {hep-th},
	reportnumber = {USTC-ICTS/PCFT-24-25},
	title = {{Correlators of N=4 Supersymmetric Yang-Mills Theory on Real Projective Space at Strong Coupling}},
	volume = {133},
	year = {2024}}

@article{Caetano:2022mus,
	archiveprefix = {arXiv},
	author = {Caetano, Joao and Rastelli, Leonardo},
	doi = {10.1007/JHEP02(2023)106},
	eprint = {2206.06375},
	journal = {JHEP},
	pages = {106},
	primaryclass = {hep-th},
	reportnumber = {CERN-TH-2022-097},
	title = {{Holography for $ \mathcal{N} $ = 4 on $ \mathbb{RP} ^{4}$}},
	volume = {02},
	year = {2023}}

@article{DeWolfe:2001pq,
	archiveprefix = {arXiv},
	author = {DeWolfe, Oliver and Freedman, Daniel Z. and Ooguri, Hirosi},
	doi = {10.1103/PhysRevD.66.025009},
	eprint = {hep-th/0111135},
	journal = {Phys. Rev. D},
	pages = {025009},
	reportnumber = {CALT-68-2359, CITUSC-01-041, NSF-ITP-01-172, MIT-CTP-3212},
	title = {{Holography and defect conformal field theories}},
	volume = {66},
	year = {2002}}

@article{Wang:2020jgh,
	archiveprefix = {arXiv},
	author = {Wang, Yifan},
	doi = {10.1007/JHEP03(2021)203},
	eprint = {2005.07197},
	journal = {JHEP},
	pages = {203},
	primaryclass = {hep-th},
	title = {{From $\mathcal{N}=4$ Super-Yang-Mills on $\mathbb{RP}^4$ to bosonic Yang-Mills on $\mathbb{RP}^2$}},
	volume = {03},
	year = {2021}}

@article{Chen:2025yxg,
    author = "Chen, Junding and Jiang, Yunfeng and Zhou, Xinan",
    title = "{Giant Graviton Correlators as Defect Systems}",
    eprint = "2503.22987",
    archivePrefix = "arXiv",
    primaryClass = "hep-th",
    reportNumber = "USTC-ICTS/PCFT-25-14",
    doi = "10.1103/hg9p-hblr",
    journal = "Phys. Rev. Lett.",
    volume = "135",
    number = "8",
    pages = "081602",
    year = "2025"
}

@article{Maldacena:1997re,
    author = "Maldacena, Juan Martin",
    title = "{The Large $N$ limit of superconformal field theories and supergravity}",
    eprint = "hep-th/9711200",
    archivePrefix = "arXiv",
    reportNumber = "HUTP-97-A097, HUTP-98-A097",
    doi = "10.4310/ATMP.1998.v2.n2.a1",
    journal = "Adv. Theor. Math. Phys.",
    volume = "2",
    pages = "231--252",
    year = "1998"
}

@article{Witten:1998qj,
    author = "Witten, Edward",
    title = "{Anti de Sitter space and holography}",
    eprint = "hep-th/9802150",
    archivePrefix = "arXiv",
    reportNumber = "IASSNS-HEP-98-15",
    doi = "10.4310/ATMP.1998.v2.n2.a2",
    journal = "Adv. Theor. Math. Phys.",
    volume = "2",
    pages = "253--291",
    year = "1998"
}

@article{Gubser:1998bc,
    author = "Gubser, S. S. and Klebanov, Igor R. and Polyakov, Alexander M.",
    title = "{Gauge theory correlators from noncritical string theory}",
    eprint = "hep-th/9802109",
    archivePrefix = "arXiv",
    reportNumber = "PUPT-1767",
    doi = "10.1016/S0370-2693(98)00377-3",
    journal = "Phys. Lett. B",
    volume = "428",
    pages = "105--114",
    year = "1998"
}

@article{Chen:2026ium,
    author = "Chen, Junding and Jiang, Yunfeng and Zhou, Xinan",
    title = "{Defect approach to giant graviton dynamics}",
    eprint = "2602.13570",
    archivePrefix = "arXiv",
    primaryClass = "hep-th",
    reportNumber = "USTC-ICTS/PCFT-26-14",
    doi = "10.1088/1361-6633/ae79a4",
    journal = "Rept. Prog. Phys.",
    volume = "89",
    number = "6",
    pages = "067801",
    year = "2026"
}

@inproceedings{Zhou:2026xtp,
    author = "Zhou, Xinan",
    title = "{Pre-Strings Lectures on Holographic Correlators and Analytic Bootstrap}",
    booktitle = "{Pre{\nobreakdash-}Strings School 2026}",
    eprint = "2609.01986",
    archivePrefix = "arXiv",
    primaryClass = "hep-th",
    month = "9",
    year = "2026"
}

@article{Caetano:2026jee,
    author = "Caetano, Joao and Komatsu, Shota and Soresina, Paolo and Wallberg, Anders Heide",
    title = "{Orientifold and Integrability: $\mathcal{N}=4$ Super-Yang-Mills on $\mathbb{RP}^4$}",
    eprint = "2610.28725",
    archivePrefix = "arXiv",
    primaryClass = "hep-th",
    month = "10",
    year = "2026"
}

@article{Chen:2026yzm,
    author = "Chen, Qi and Huang, Zhongjie and Jiang, Yunfeng and Zhou, Xinan",
    title = "{Giant Graviton Correlators: From Non-maximal to Dual, Bootstrap and Hidden Symmetry}",
    eprint = "2611.13444",
    archivePrefix = "arXiv",
    primaryClass = "hep-th",
    month = "11",
    year = "2026"
}
\bibliographystyle{utphys}
\end{document}